\documentclass[usedcolumn,fleqn,usenatbib]{mnras}

\usepackage{newtxtext,newtxmath}

\usepackage[T1]{fontenc}
\usepackage{ae,aecompl}

\usepackage{graphicx}	
\usepackage{amssymb}	
\usepackage{color}

\newcommand{\kepler}{{\it Kepler}}
\newcommand{\corot}{{\it CoRoT}}
\newcommand{\tess}{{\it TESS}}
\newcommand{\plato}{{\it PLATO}}
\newcommand{\gaia}{{\it Gaia}}
\newcommand{\jwst}{{\it JWST}}
\newcommand{\gpe}{GP-EBOP}
\newcommand{\ms}{\ensuremath{\rm m\,s^{-1}}}
\newcommand{\kms}{\ensuremath{\rm km\,s^{-1}}}
\newcommand{\casu}{{\sc CASUTools}}

\title[Next Generation Transit Survey]{\vspace{-0.5cm}The Next Generation Transit Survey (NGTS)
\vspace{-0.5cm}
}

\author[P.\ J.\ Wheatley et al.]{
\parbox{\textwidth}{
Peter~J.~Wheatley,$^{1,2}$\thanks{E-mail: P.J.Wheatley@warwick.ac.uk}
Richard~G.~West,$^{1,2}$
Michael~R.~Goad,$^{3}$
James~S.~Jenkins,$^{4,5}$
Don~L.~Pollacco,$^{1,2}$
Didier~Queloz,$^{6}$
Heike~Rauer,$^{7,8}$
St\'{e}phane~Udry,$^{9}$\\
Christopher~A.~Watson,$^{10}$
Bruno Chazelas,$^{9}$
Philipp~Eigm\"uller,$^{7}$
Gregory Lambert,$^{6}$\\
Ludovic Genolet,$^{9}$
James~McCormac,$^{1,2}$
Simon~Walker,$^{1}$
David~J.~Armstrong,$^{1,2}$
Daniel~Bayliss,$^{9}$
Joao~Bento,$^{1,11}$
Fran\c{c}ois Bouchy,$^{9}$
Matthew~R.~Burleigh,$^{3}$
Juan~Cabrera,$^{7}$
Sarah~L.~Casewell,$^{3}$
Alexander~Chaushev,$^{3}$
Paul~Chote,$^{1}$\\
Szil\'ard~Csizmadia,$^{7}$
Anders~Erikson,$^{7}$
Francesca~Faedi,$^{1}$
Emma~Foxell,$^{1,2}$\\
Boris~T.~G\"ansicke,$^{1,2}$
Edward~Gillen,$^{6}$
Andrew Grange,$^{3}$
Maximilian~N.~G{\"u}nther,$^{6}$
Simon~T.~Hodgkin,$^{11}$
James~Jackman,$^{1,2}$
Andr\'{e}s~Jord\'{a}n,$^{13,14,15}$
Tom~Louden,$^{1,2}$\\
Lionel Metrailler,$^{9}$
Maximiliano~Moyano,$^{16}$
Louise~D.~Nielsen,$^{9}$
Hugh~P.~Osborn,$^{1}$\\
Katja~Poppenhaeger,$^{10}$
Roberto~Raddi,$^{1}$
Liam~Raynard,$^{3}$
Alexis~M.~S.~Smith,$^{7}$\\
Maritza~Soto$^{4}$,
Ruth~Titz-Weider$^{7}$
}
\vspace{0.1cm}
\\
$^{1}$Dept.\ of Physics, University of Warwick, Gibbet Hill Road, Coventry CV4 7AL, UK\\
$^{2}$Centre for Exoplanets and Habitability, University of Warwick, Gibbet Hill Road, Coventry CV4 7AL, UK\\
$^{3}$Dept.\ of Physics and Astronomy, University of Leicester, University Road, Leicester, LE1 7RH, UK\\
$^{4}$Departamento de Astronomia, Universidad de Chile, Casilla 36-D, Santiago, Chile\\
$^{5}$Centro de Astrof\'isica y Tecnolog\'ias Afines (CATA), Casilla 36-D, Santiago, Chile.\\
$^{6}$Astrophysics Group, Cavendish Laboratory, J.J. Thomson Avenue, Cambridge CB3 0HE, UK \\
$^{7}$Institute of Planetary Research, German Aerospace Center, Rutherfordstrasse 2, 12489 Berlin, Germany\\
$^{8}$Center for Astronomy and Astrophysics, TU Berlin, Hardenbergstr. 36, D-10623 Berlin, Germany\\
$^{9}$Observatoire Astronomique de l'Universit\'{e} de Gen\`{e}ve, 51 Ch. des Maillettes, 1290 Versoix, Switzerland\\
$^{10}$Astrophysics Research Centre, School of Mathematics and Physics, Queen's University Belfast, BT7 1NN Belfast, UK\\
$^{11}$Research School of Astronomy and Astrophysics, Mount Stromlo Observatory, Australian National University, Cotter Road, Weston, ACT 2611, Australia\\
$^{12}$Institute of Astronomy, University of Cambridge, Madingley Rise, Cambridge CB3 0HA, UK\\
$^{13}$Instituto de Astrof\'{i}sica, Facultad de F\'{i}sica, Pontificia Universidad Cat\'{o}lica de Chile, Av. Vicu\~{n}a Mackenna 4860, 7820436 Macul, Santiago, Chile\\
$^{14}$Max-Planck-Institut f\"ur Astronomie, K\"onigstuhl 17, 69117 Heidelberg, Germany\\
$^{15}$Millennium Institute of Astrophysics, Santiago, Chile\\
$^{16}$Instituto de Astronomia, Universidad Cat\'{o}lica del Norte, Casa Central, Angamos 0610, Antofagasta, Chile\\
\vspace{-0.9cm}
}

\date{Accepted 2017 October 28. Received 2017 October 27; in original form 2017 September 13\vspace{-0.4cm}}

\pubyear{2017}

\begin{document}
\label{firstpage}
\pagerange{\pageref{firstpage}--\pageref{lastpage}}
\maketitle

\begin{abstract}
We describe the Next Generation Transit Survey (NGTS), which is a ground-based project searching for transiting exoplanets orbiting bright stars. NGTS builds on the legacy of previous surveys, most notably WASP, and is designed to achieve higher photometric precision and hence find smaller planets than have previously been detected from the ground. It also operates in red light, maximising sensitivity to late K and early M dwarf stars. The survey specifications call for photometric precision of 0.1 per cent in red light over an instantaneous field of view of 100 square degrees, enabling the detection of Neptune-sized exoplanets around Sun-like stars and super-Earths around M dwarfs. The survey is carried out with a purpose-built facility at Cerro Paranal, Chile, which is the premier site of the European Southern Observatory (ESO). An array of twelve 20\,cm f/2.8 telescopes fitted with back-illuminated deep-depletion CCD cameras are used to survey fields intensively at intermediate Galactic latitudes. The instrument is also ideally suited to ground-based photometric follow-up of exoplanet candidates from space telescopes such as \tess, \gaia\ and \plato. We present observations that combine precise autoguiding and the superb observing conditions at Paranal to provide routine photometric precision of 0.1 per cent in 1 hour for stars with I-band magnitudes brighter than 13. We describe the instrument and data analysis methods as well as the status of the survey, which achieved first light in 2015 and began full survey operations in 2016. 
NGTS data will be made publicly available through the ESO archive.
\end{abstract}

\vspace{-1cm}


\section{Introduction}
\label{sec-intro}
The photometric detection of transits has proved to be the key to determining a wide range of the physical characteristics of exoplanets. The depth of a transit depends on the relative radii of planet and star ($R_{\rm p}/R_{\rm *}$) and the first transit detections immediately showed that hot Jupiters are gas giants and not composed primarily of heavy elements \citep{Henry00, Charbonneau00}. Transits also enabled the measurement of stellar obliquities using the Rossiter-McLaughlin effect \citep{Winn05}, with important implications for exoplanet migration \citep[e.g.][]{Triaud10,Albrecht12}. They also present the opportunity to determine the composition and structure of planetary atmospheres through transmission spectroscopy \citep[e.g.][]{Charbonneau02,Sing16}, with detections of Doppler shifts revealing planetary winds \citep{Snellen10, Louden15} and the detection of deep transits in ultraviolet lines revealing planetary evaporation \citep[e.g.][]{Vidal03,Ehrenreich15}. Detections of secondary eclipses and phase curves in transiting systems allow determination of the reflected and thermal emission spectra of exoplanets, together with albedos and the efficiency of heat transport around the planet \citep[e.g.][]{Deming05,Charbonneau05,Knutson07}. When coupled with mass determinations based on the radial-velocities of the star, transits also provide planetary densities and hence constraints on their bulk composition and internal structure \citep[e.g.][]{Seager07,Baraffe08}.

\begin{figure}
	\includegraphics[width=8.4cm]{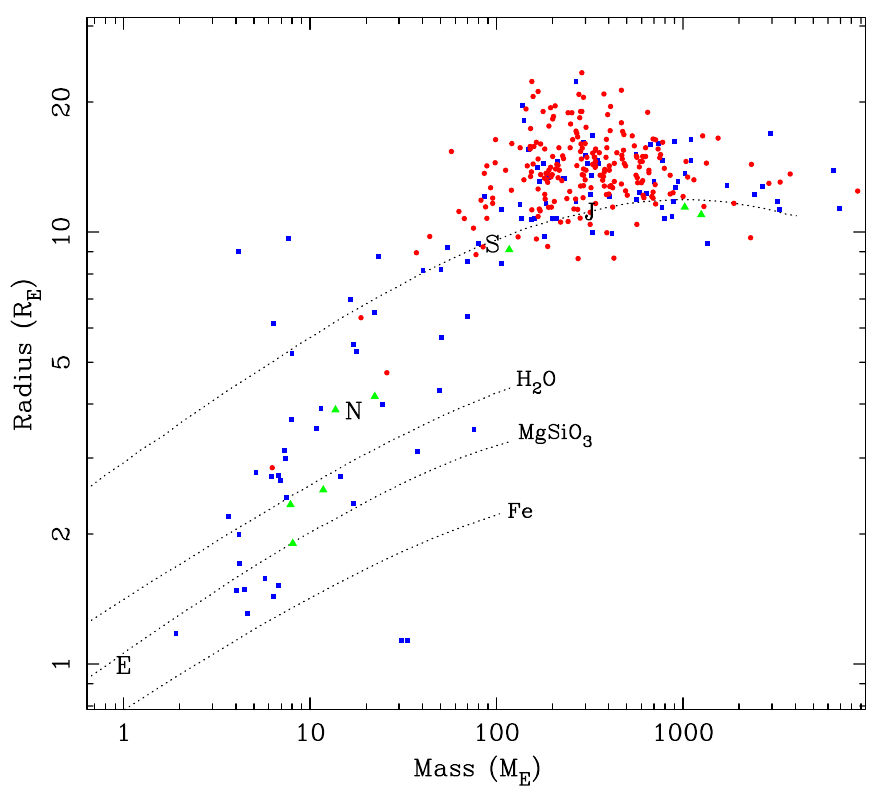}
    \caption{The mass-radius relation for known transiting exoplanets with masses determined to better than 20 per cent precision (taken from the NASA Exoplanet Archive in January 2017). Planets initially discovered in ground-based transit surveys are plotted as red circles, while those detected from space are plotted as blue squares, and planets initially found from radial velocity measurements are plotted as green triangles. Solar system planets are indicated with letters and the dotted lines are mass-radius relations calculated for different compositions by \citet{Seager07}. }
    \label{fig-mr}
\end{figure}

\begin{figure}
	\includegraphics[width=8.4cm]{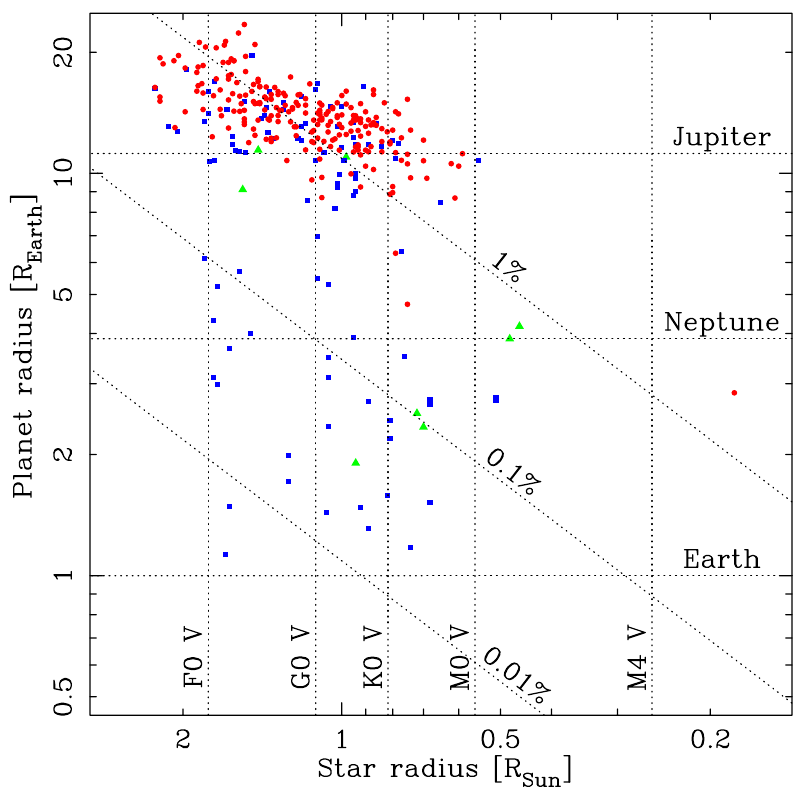}
    \caption{The planet and star radii for known transiting exoplanets with masses determined to better than 20 per cent precision. Diagonal dotted lines indicate systems with equal transit depth, while the horizontal lines show the radii of solar system planets and the vertical lines are indicative of stellar spectra types \citep[Mamajek, priv.\ comm., based on][]{Pecaut13}. The colours and symbols are the same as Fig.\,\ref{fig-mr}.
    }
    \label{fig-rr}
\end{figure}

A prerequisite for the application of 
this wide range of powerful techniques
in exoplanet characterisation
is the discovery of transiting exoplanets, usually in wide-field photometric surveys. Since most of the characterisation methods require high signal-to-noise
measurements, there is particular value in the detection of transiting planets around bright stars.

The most successful ground-based surveys for transiting exoplanets have been WASP \citep{Pollacco06}, HATNet \citep{Bakos04} and HATSouth \citep{HATS}, which together account for more than 50 per cent of all the known transiting planets with masses determined to better than 20 per cent (including those found from space).
WASP and HATNet employ telephoto lenses mounted on CCD cameras to make precise photometric measurements over large swaths of the sky, while HATSouth employs 
24 telescope tubes spread over three locations in the southern hemisphere.
Typically these surveys have found planets around the mass of Saturn to a few times the mass of Jupiter, and with radii between that of Saturn and twice Jupiter (Fig.\,\ref{fig-mr}). A handful of smaller transiting exoplanets have also been found
in ground-based transit surveys
\citep{Charbonneau09,Bakos10,Berta-Thompson15,Gillon16,Gillon17,Dittmann17}
and transits have been found for some planets initially identified in ground-based radial velocity surveys
\citep{Gillon07,Winn11,Bonfils12,Dragomir13,Motalebi15}.
The full population of transiting exoplanets with masses determined to better than 20 per cent is shown in Fig.\,\ref{fig-mr} (sample taken from the NASA Exoplanet Archive\footnote{\url{http://exoplanetarchive.ipac.caltech.edu/}} in January 2017). The figure illustrates the remarkably diverse nature of the known population, including for instance a factor of eight range in density of Jupiter-mass planets. The heating mechanism that inflates the low density hot Jupiters remains a matter of debate \citep[e.g.][]{Spiegel13}.

Space-based surveys, most notably \kepler\ \citep{Borucki10} and \corot\ \citep{Auvergne09}, have made more precise photometric measurements and have thereby discovered transiting exoplanets with smaller radii. These have included rocky exoplanets \citep[e.g.][]{Leger09,Queloz09,Batalha11}, multi-planet systems \citep[e.g.][]{Lissauer11} and even circumbinary planets \citep[e.g.][]{Doyle11,Welsh12}. Thousands of candidates have been identified, although to date the space-based surveys have covered a relatively small proportion of the sky. As a consequence, most of the detected candidates are too faint for radial-velocity confirmation and mass determination, and their masses remain poorly constrained. For some multi-planet systems it is possible to use Transit Timing Variations (TTVs) to place constraints on planet masses \citep[e.g.][]{Lissauer13}, but the mass-radius relation remains relatively sparsely populated below the mass of Saturn (see Fig.\,\ref{fig-mr}).

Transits of Earths and super-Earths around Sun-like stars have very shallow depths that are currently only detectable from space, and the discovery of new examples around bright stars depends on the extended \kepler\ K2 mission, and
new missions such as \tess\ \citep{Ricker15} and \plato\ \citep{Rauer14}.
In contrast, Neptunes around Sun-like stars and Earths and super-Earths around late-type dwarfs have transit depths that should be detectable in ground-based surveys. Current examples include the super-Neptune HAT-P-11b \citep{Bakos10}, the super-Earths GJ1214b, GJ1132b and LHS\,1140b
\citep{Charbonneau09,Berta-Thompson15,Dittmann17}, and the Earths around TRAPPIST-1
\citep{Gillon16,Gillon17}.
This region of parameter space is important because it includes the transitions between gas and ice giants and between volatile-rich and volatile-poor super-Earths. The 
population of systems that have been well-characterised to date (Fig.\,\ref{fig-mr}) indicate that these smaller planets exhibit a diversity that is at least comparable to that of the gas giants.

In this paper we describe a new ground-based instrument
that has been designed to discover new transiting exoplanets in these size ranges and to follow up candidate exoplanets from space telescopes: the Next Generation Transit Survey (NGTS).

\begin{figure}
	\includegraphics[width=8.4cm]{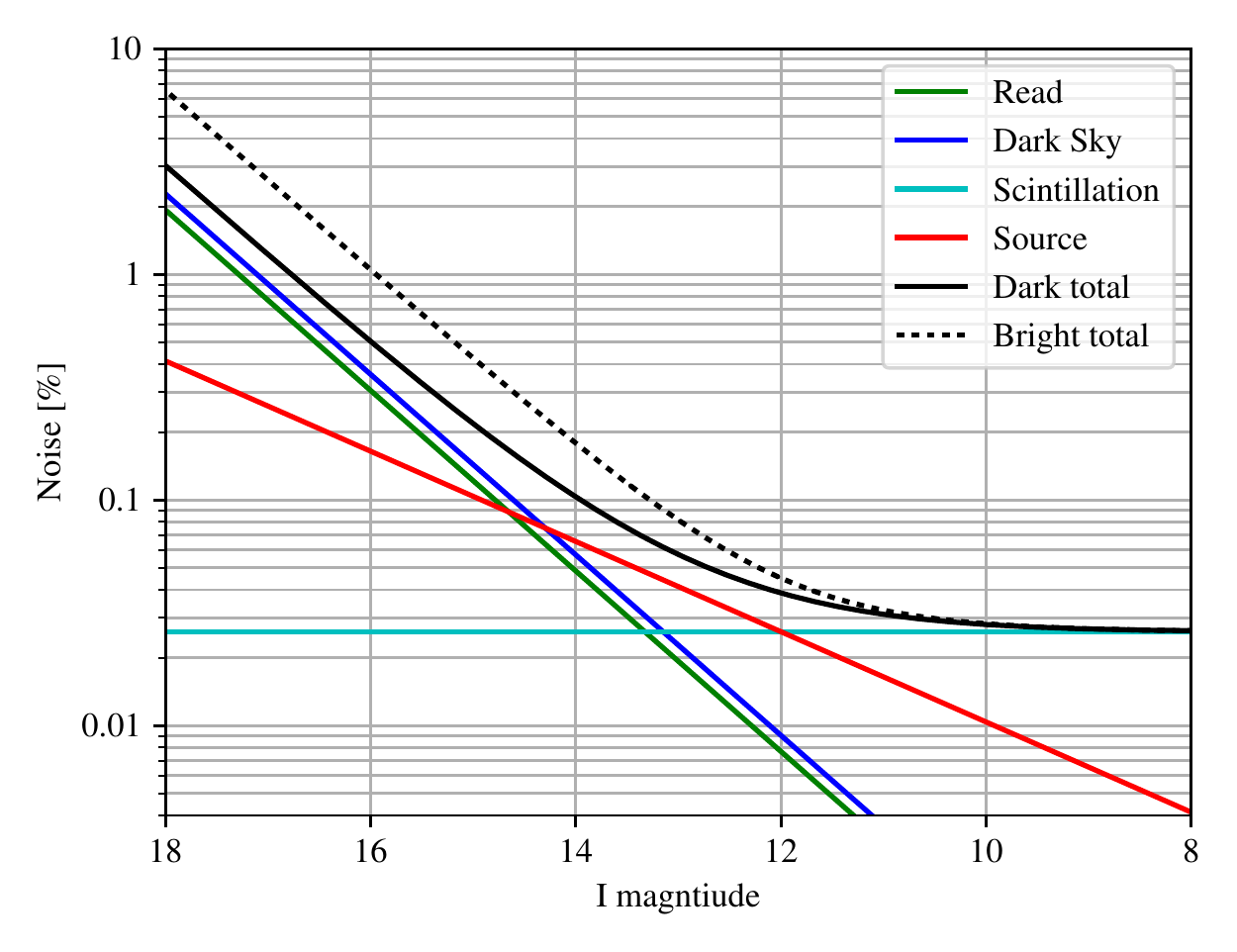}
    \caption{A schematic noise model for 1\,h exposure time with one of the NGTS telescopes at Paranal, which have 20\,cm aperture and pixel scale of 5\,arcsec. The solid black line is the best-case total noise for dark sky, and the dotted line shows the noise for full Moon. Poisson noise from background light dominates for fainter stars, with Poisson noise from the target star being significant at intermediate brightness, and atmospheric scintillation dominating for bright stars. Detector read noise begins to become significant with dark sky (for individual 10\,s exposures) but dark current is negligible for a sufficiently cooled high quality device.
    }
    \label{fig-noise}
\end{figure}

\section{Science goals and design considerations}
\label{sec-goals}
The primary science goal of NGTS is to extend the wide-field ground-based detection of transiting exoplanets to at least the Neptune size range, in particular for stars that are sufficiently bright for radial-velocity confirmation and mass determination. This will allow us to determine the density of these exoplanets, and hence their bulk composition,  better populating the exoplanet parameter space shown in Fig.\,\ref{fig-mr}. These exoplanets will also be suitable targets for the wide range of characterisation techniques outlined in Sect.\,\ref{sec-intro}, including studies of their atmospheric structure and composition. 

The NGTS facility will also enable efficient photometric follow up of transit candidates identified in space-based transit surveys. 
Since most of the sky is visible to NGTS for much longer than the 27\,d dwell time of the \tess\ survey \citep{Ricker15}, NGTS can measure the orbital periods of exoplanets detected with single transits by \tess. 
NGTS will also
test transit candidates by searching for blended variable stars with a finer pixel scale than either \tess\ or \plato\ \citep{Rauer14}
and it will refine ephemerides of key exoplanets in advance of observations with flagship facilities such as \jwst\ and E-ELT.
NGTS will also search for transits of exoplanets detected by the radial velocity and astrometry methods \citep[e.g.\ with \gaia;][]{Perryman14}.

In Fig.\,\ref{fig-rr} we illustrate the parameter space relevant to transit detection, plotting the radius of known transiting exoplanets against the radius of the host star. In this diagram the limiting transit depth of a survey corresponds to a diagonal line. It can be seen that most transiting planets identified in ground-based surveys have transit depths around 1 per cent, but there are a few with significantly shallower transits. Careful follow-up observations of transits with ground-based instruments often achieves sub-mmag precision \citep[$<$0.1\,per cent; e.g.][]{Southworth09,Kirk16} and optical secondary eclipses of hot Jupiters with depths around 0.1 per cent have also been detected from the ground \citep[e.g.][]{Sing09,Burton12}. It should therefore be possible, at least in principle, to design a ground-based instrument that can achieve similar precision over a wide field of view.

In order to be sensitive to Neptune-sized planets around Sun-like stars, and sub-Neptunes around later type stars, we set ourselves the goal of detecting 0.1 per cent depth transits. As can be seen in Fig.\,\ref{fig-rr}, this precision would correspond to super-Earths around early M dwarfs. In order to populate the mass-radius relation of exoplanets in these size ranges it is crucial that the target stars are sufficiently bright for radial velocity confirmation and precise mass determination. This sets faint visual magnitude limits of around 13 and 15 respectively for HARPS and ESPRESSO \citep{Mayor03,Pepe14}. Populating parameter space also requires a statistical sample of such planets, and Monte Carlo simulations using the Kepler occurrence rates show that an instrument with an instantaneous field of view of around $100\,\rm deg^2$ is needed in order to detect a sample of tens of small planets in a survey lasting a few years \citep{Wheatley13,Guenther17}.

\begin{figure}
	\includegraphics[width=8.4cm]{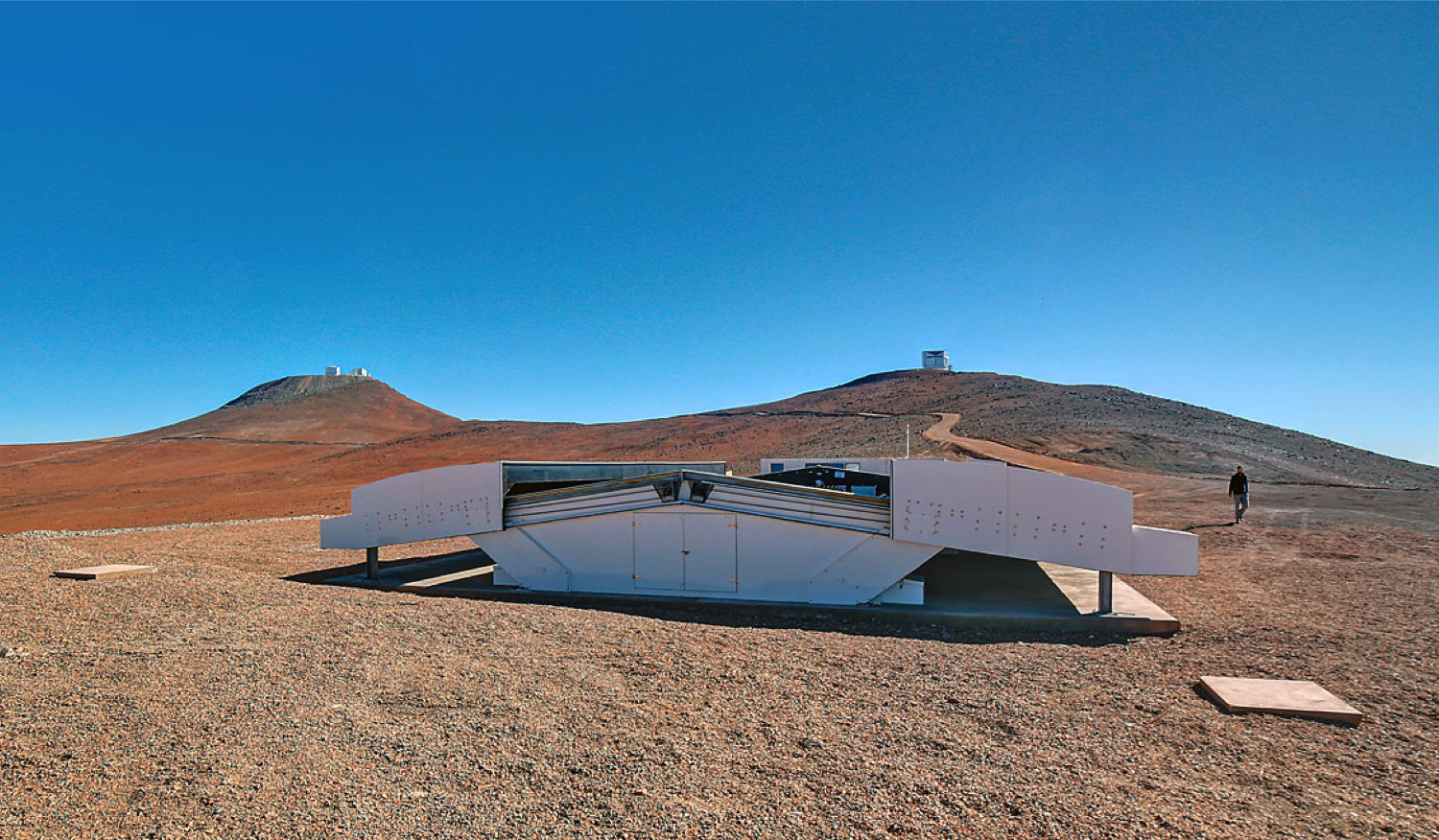}
    \caption{The NGTS enclosure at the ESO Paranal observatory, Chile. The ESO Very Large Telescope (VLT) and VISTA telescope can be seen in the background (left and right respectively).}
    \label{fig-ngts}
\end{figure}

In contrast to space-based observing, ground-based telescopes have to contend with a number of atmospheric effects that act to degrade photometric precision. These include extinction, which varies with time and telescope elevation \citep[e.g.][for Cerro Paranal]{Noll12}; increased and variable sky background due to airglow and scattered light from the Sun and Moon \citep[also][]{Noll12}; and scintillation due to atmospheric turbulence \citep[e.g.][]{Osborn15}.
In principle, atmospheric extinction can be fully modelled and eliminated from the noise budget, although a superb site is required to provide reliably clear skies and low and stable extinction. Sky background and atmospheric scintillation, on the other hand, do limit the precision of ground-based photometry. This is illustrated in Fig.\,\ref{fig-noise} where we plot a schematic noise model for an exposure time of one hour with the baseline design of NGTS: a 20\,cm telescope equipped with 5\,arcsec pixels based at the ESO Paranal Observatory in Chile (assuming also a 3 pixel radius aperture and individual exposures of 10\,s). For bright stars the precision in this model is limited by atmospheric scintillation \citep[for which we have adopted the scaling law for Paranal from][]{Osborn15}, for intermediate brightness stars the precision is limited by Poisson noise from the target star, and for faint stars it is limited by Poisson noise from background light (with a contribution during dark time also from read noise, which is assumed here to be 15 electrons). Nevertheless, it can be seen that 0.1\,per cent photometry on the timescale of a single transit is possible for stars brighter than I=14 in dark time and for stars brighter than I=13 even at full Moon.
This demonstrates that a large telescope aperture is not needed to achieve our science goals.
Although, of course, this  model does not account for additional noise associated with the instrument (e.g.\ non-linearity or flat fielding). In principle these additional noise sources can be eliminated, but in practice they tend to be significant and require careful treatment.

\begin{figure}
	\includegraphics[width=8.4cm]{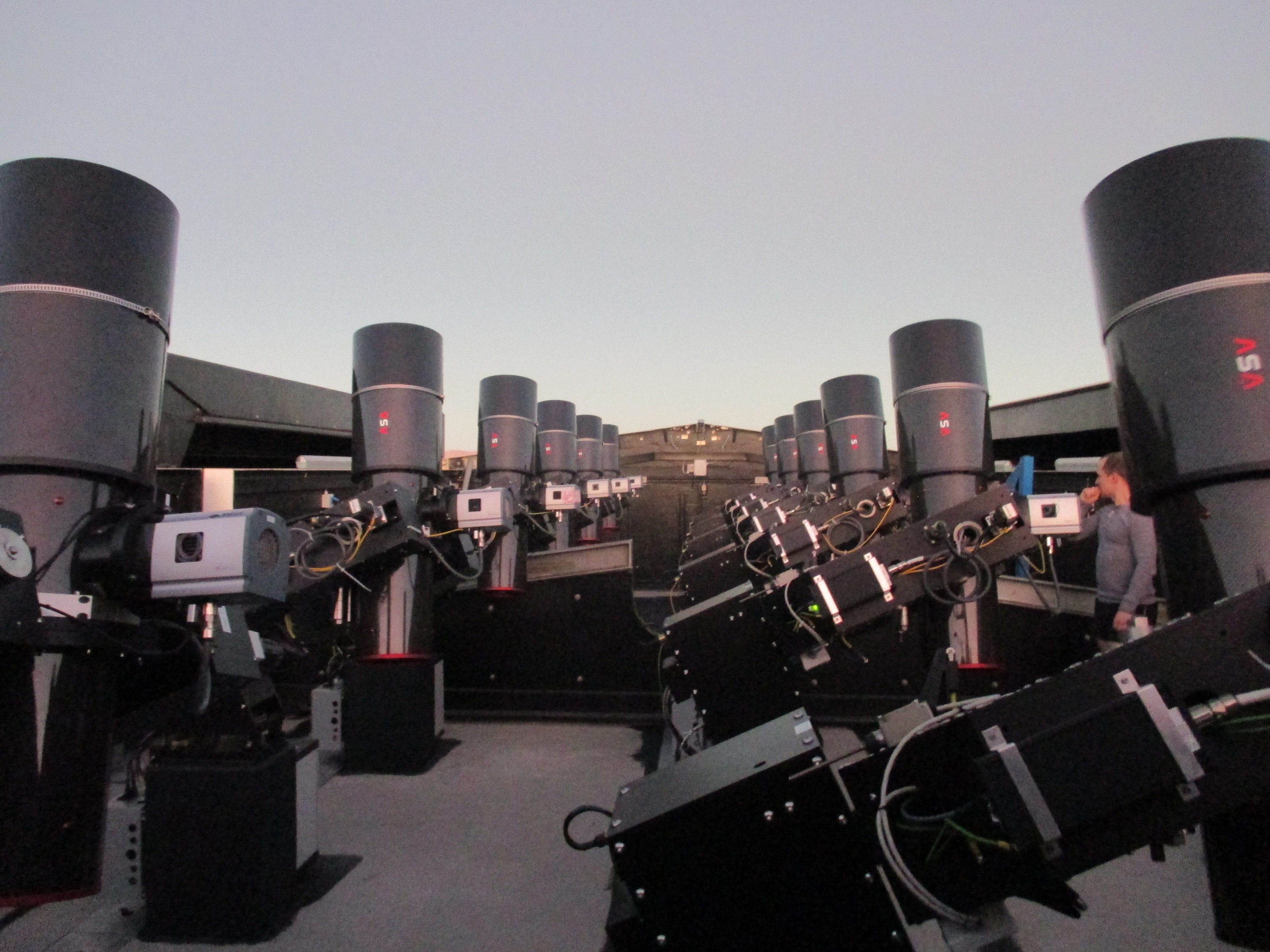}
    \caption{The twelve 20\,cm telescopes of the NGTS facility taking flat field images in evening twilight at Cerro Paranal, Chile.}
    \label{fig-tels}
\end{figure}

Another important effect for wide-field ground-based telescopes is atmospheric refraction, which acts to stretch the field of view of a telescope at lower elevations. This limits the size of field of view over which it is possible to achieve precise autoguiding (whereby stars are maintained at fixed positions on the detector). Precise autoguiding is probably needed in order to limit flat-field noise, and this was a key factor in the success of the primary \kepler\ mission 
\citep[e.g.][]{Koch10}. The alternative approach of defocusing, which is often used for transit studies of individual stars, is not an attractive option for a wide field survey since it is wasteful of pixels, which are usually the cost limiting factor. A single telescope with a field of view of $100\,\rm deg^2$ (i.e.\ $10^\circ$ across) would see stars on opposite sides of the image move apart by 24\,arcsec between the zenith and an elevation of $30^\circ$. The pixel scale of the instrument would either resolve this movement and risk flat-field noise, or would have large pixels and suffer from increased Poisson noise from background light. It is advantageous, therefore, to build the large field of view for NGTS from an array of individual telescopes, each equipped with its own independently steerable mount. This arrangement has the added advantage of versatility, allowing for efficient follow up of multiple transit candidates from space-based surveys, while including the option to maximise collecting area and photometric precision by pointing all the telescopes to the same target.

The fastest commercially available and cost effective small telescopes have focal ratios of around f/2.8, which sets the plate scale and hence field of view of each telescope unit for a given aperture and detector size. For a 20\,cm f/2.8 telescope with a 28\,mm detector, the field of view is $2.8^\circ$, and an array of 12 telescopes is needed to cover a total instantaneous field of $96\,\rm deg^2$.
A telescope with aperture larger than 20\,cm (with the same focal ratio and detector) would reduce Poisson noise by collecting more target photons and by better resolving the sky background with smaller pixels, and it would also improve scintillation noise, which scales approximately as $D^{-2/3}$, where $D$ is the diameter of the telescope aperture \citep{Young67,Dravins98}. However, these improvements would come at the cost of reduced field of view per telescope unit, increasing the number of telescope units required, and increasing the financial cost of the project. 
The number of required telescope units scales as the square of the aperture, so an aperture of 40\,cm would require 48 telescope units to cover the same field of view as our baseline 12 units. We note that our chosen combination of focal ratio and aperture is similar to that made by the HATSouth survey \citep{HATS}, who cover a similar total field of view with 8 telescope tubes from each of their three sites, although in their setup groups of 4 telescope tubes share a single mount.

In order to achieve the high photometric precision needed to detect small exoplanets it is also important for the detector to be back illuminated, in order to avoid sub-pixel sensitivity variations associated with electrode structures. There is also a significant advantage in using deep depletion CCDs, which achieve high quantum efficiency in the red optical with minimal fringing. Operating in the red optical provides maximum sensitivity to K and M dwarf stars, where smaller planets can be detected for a given limiting precision (Fig.\ref{fig-rr}), without the much higher costs and inferior photometric performance of infra-red detectors.  A survey of the available back-illuminated deep-depletion CCDs indicated that detectors with around 4\,Mpix and physical sizes of around 28\,mm provided the most cost-effective solution.

\section{The NGTS facility}
\label{sec-ngts}
The NGTS instrument
was constructed during 2014 and 2015 at the ESO Paranal Observatory in northern Chile. It consists of an array of twelve independently-steerable 20\,cm telescopes, with a combined instantaneous field of view of $96\,\rm deg^2$. The telescopes are housed in a single enclosure, with a roll-off roof, located about 900\,m from ESO's VISTA telescope and at an altitude of 2440\,m. 
The site and enclosure are shown in Fig.\,\ref{fig-ngts} and the telescope units are shown in
Fig.\,\ref{fig-tels}.
Aspects of the design of the survey were described by \citet{Chazelas12}, and a prototype system demonstrating some of the key technologies was tested on La Palma in 2010, with results presented by \citet{McCormac17}.

\subsection{Telescopes}
\label{sec-tels}
\begin{figure}
	\includegraphics[width=8.4cm]{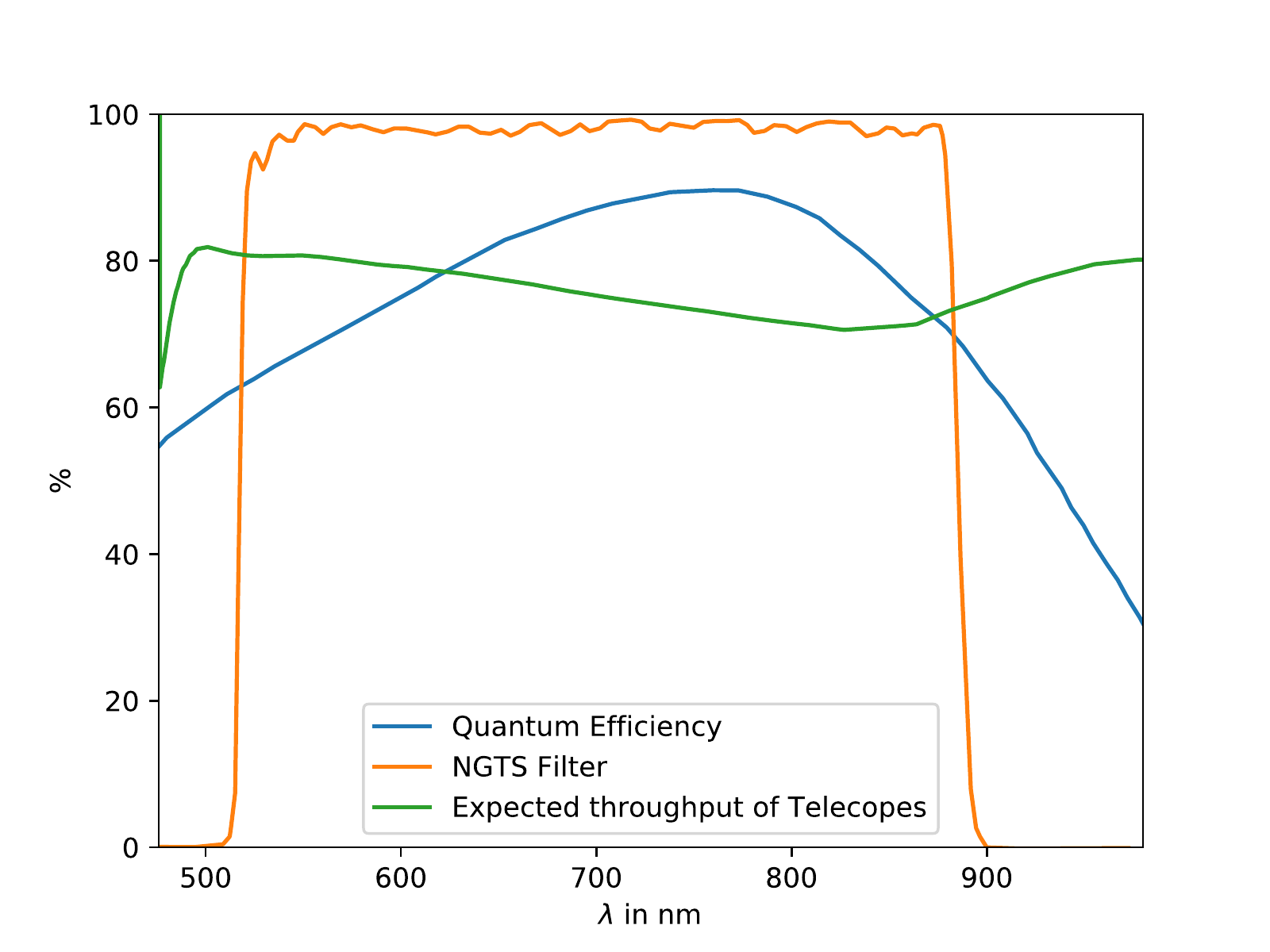}
    \caption{Theoretical throughput curve for the NGTS telescope (green), the measured throughput of the NGTS filter (orange), and the expected quantum efficiency of the detector (blue). A data file containing the combined throughput curve of NGTS is available online.}
    \label{fig-qe}
\end{figure}

The NGTS telescopes are a custom version of the  8\,inch f/2.8 H Astrograph supplied by AstroSysteme Austria\footnote{\url{http://www.astrosysteme.com}} (ASA). This a Newtonian design with a 20\,cm hyperbolic primary mirror constructed from Suprax (with low thermal expansion), a flat 9\,cm secondary mirror (giving 20\,per cent obscuration) and a four-element corrector lens. Vignetting is 9\,per cent at the edge of the field of view. The mirror is coated with protected aluminium, which does not provide the best reflectivity in the red optical, but was selected for its stability in the Paranal environment. The expected throughput of the telescope is plotted in Fig.\,\ref{fig-qe}. The mirror is further protected by a window of optical-quality 8\,mm BK7 glass (wave front error of $\sim\lambda/4$\, RMS) that has been installed at the top of the telescope tube, preventing dust and other contaminants falling directly onto the primary mirror.
Our corrector lens is of a custom design that ensures a point spread function (PSF) of below $12\rm\,\mu m$ ($<$1\,pixel) across a field of 38\,mm diameter (larger than the detector). The corrector lenses and the protective windows have anti-reflection coatings optimised for the wavelength range 500--1000\,nm, which spans the NGTS filter bandpass. The telescope optics are collimated on-site with a laser alignment tool.

The NGTS filter has been specifically designed for the experiment with a bandpass from 520 to 890\,nm, which provides good sensitivity to late K and early M dwarfs. The red cut-off is designed to minimise variations in the atmospheric extinction by excluding the strong water absorption bands beyond 900\,nm, which are highly variable even at Paranal \citep{Noll12}. While slightly wasteful of the red sensitivity of the detector (Sect.\,\ref{sec-cameras} and Fig.\,\ref{fig-qe}), this choice ensures that the effective bandpass of NGTS is defined primarily by the instrument rather than the sky. The measured throughput of the filter is plotted in Fig.\,\ref{fig-qe} and a data file with the combined throughput of the NGTS system is available with the online version of this article.

The telescope tubes are made from carbon fibre and the mechanical design has been customised with a metal ring in order to interface with the telescope mounts that were supplied by a different manufacturer (Sect.\,\ref{sec-mounts}).  
As the Newtonian configuration is particularly susceptible to scattered light, it has been necessary to install baffles of 600 mm in two sections in front of the telescope tube (which are visible in Fig.\,\ref{fig-tels}). These baffles ensure that at $30^\circ$ from the moon there is no direct illumination of the protective glass installed at the entrance of the telescope tube.

The telescopes are fitted with electronic focusers that are controlled using a serial connection from our data acquisition computers (Sect.\,\ref{sec-facility}). The mass of the NGTS camera  is at the upper end of the specification of the focuser, 
and this necessitated customisation of the mechanical components of the focuser in order to prevent excessive wear leading to play in the focus set point. The carbon fibre tube and low expansion mirrors provide good focus stability with temperature, and it is not necessary to refocus the telescope during the night. 

\subsection{Cameras}
\label{sec-cameras}
Each of the twelve NGTS telescopes is fitted with a $2048\times2048$\, pixel
CCD manufactured by e2v technologies plc\footnote{\url{http://www.e2v.com}}.
These are the deep depletion version of the CCD42-40 back-illuminated CCD sensor, which provides excellent quantum efficiency in the red optical with minimal fringing.
The quantum efficiency is plotted in Fig.\,\ref{fig-qe} and it is included in the calculation of total throughput available as an online data file.
The device has 13.5\,$\rm \mu m$ pixels and an image area of $27.6\times27.6$\,mm. The deep depletion version has higher dark current than the standard device, because it employs non-inverted mode operation (NIMO), however the dark current is negligible in our application since it remains much lower than the sky background.

The CCDs are packaged into cameras by Andor Technology Ltd\footnote{\url{http://www.andor.com}}. The camera is a custom version of the Andor iKon-L 936 camera, which is an updated version of the cameras used for the WASP project \citep{Pollacco06}. The NGTS cameras have a 4-stage thermo-electric cooler, allowing us to operate at a CCD temperature of $\rm -70^\circ C$ for all ambient conditions at Paranal.
The cameras have a custom CCD window, which is optimised for 500--950\,nm, and a custom shutter with an aperture of 45\,mm in order to accommodate the fast f/2.8 beam from the telescope (this shutter is now fitted to the iKon-L as standard). The cameras also have a custom faceplate designed to interface with the telescope focuser unit, with an O-ring to prevent dust ingress. The CCDs are aligned to the focal plane of the telescope to a precision of $0.04^\circ$ in an iterative process in which focus gradients in sky images are eliminated by adding and removing shims on the mounting bolts. We read the CCDs at a speed of 3\,MHz, reading out an entire image in 1.5\,s with a read noise of around 12\,electrons. 
We have the option to reduce read noise by reading the CCDs at 1\,MHz, but at the cost of longer readout time (4.5\,s) and reduced on-sky exposure time. 


We have worked in collaboration with Andor to make a number of modifications to the standard camera with the goal of optimising photometric precision. These modifications have included changes to the analogue readout electronics to maximise bias and gain stability, optimisation of the collection phase CCD voltage to maximise charge conservation for saturated stars, and control of the internal cooling fan for thermal stability.

Communication with the camera is via a USB connection, which is carried to our data acquisition computers over optical fibre (Sect.\,\ref{sec-facility}).
The power supply is external to the camera and provides stabilised voltages via a shielded cable. The power supply has been mounted on one of the forks of the telescope mount in order to minimise mechanical stress on the cable. Images with a prototype telescope unit suffered pick up noise originating in the telescope mount power supply, and this was eliminated by additional cross bonding of all components.

\subsubsection{Laboratory characterisation of cameras}
\label{sec-lab}
Each of our thirteen CCD cameras (including one spare) were characterised in a laboratory of the Space Research Centre at the University of Leicester before shipping to Paranal. Our goal was to characterise the cosmetic and noise properties of the cameras, with a particular focus on the wavelength dependence of the flat field, which is difficult to measure with on-sky observations. 

\begin{figure}
	\includegraphics[height=3.28cm]{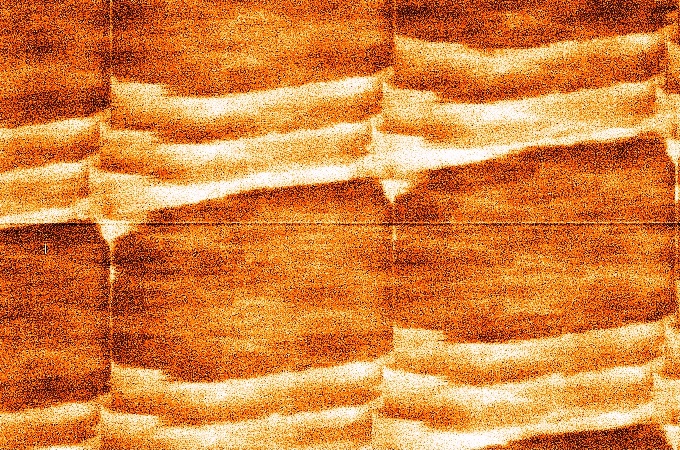}
	\includegraphics[height=3.28cm]{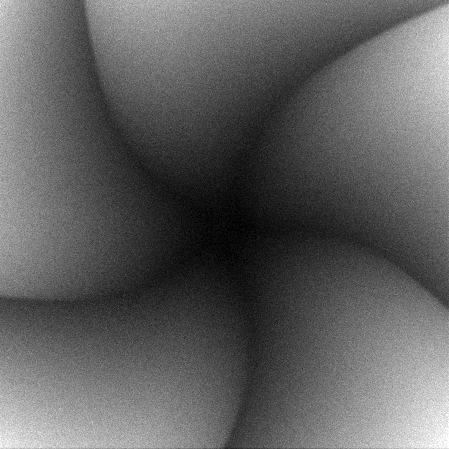}   
    \caption{Example images from our programme of laboratory characterisation of the NGTS cameras. 
Left: A small portion of an image (8 percent of the full frame) showing the ratio of laboratory flat fields measured at wavelengths of 450 and 880\,nm. The {\em blue diamond} effect can be seen, whereby the blue sensitivity of the CCD  varies spatially with an amplitude of around 2 per cent.
Right: An example shutter image, which shows the relative exposure times in different parts of the image due to the opening/closing time of the shutter petals. 
}
    \label{fig-lab}
\end{figure}

The cameras were mounted in turn on a computer-controlled movable stage, 
attached to an optical bench, 
allowing small independent movements in the x, and y
direction. They were illuminated with a electroluminescent panel (ELP), and for flat field measurements a lens and narrow band filter were used to focus a beam of known wavelength onto the CCD. The moveable stage was used to position the focussed beam onto a 9x9 grid covering the whole imaging area, with overlaps of around 200\,pixels. The master flat field frames for each wavelength were constructed by combining many images at each position and deconvolving the CCD response from the illumination function. We constructed master flats at wavelengths of 450, 650 and 880\,nm, each of which contains $>5$ million counts per pixel.

Inspection of the master flat field images showed that the CCDs are of very high cosmetic quality with no bad columns and only a handful of pixels with low sensitivity.
These bad pixels were mapped so that affected on-sky measurements of stars could be flagged in the pipeline reduction, and a confidence map was constructed in order to down-weight pixels with poor or unstable response.  As expected, the flat fields also exhibited a wavelength dependence in a regular pattern, on a scale of 10s to 100s of pixels, that we understand to be related to the manufacturing process by which the back illuminated chips are thinned (the so-called {\em blue diamond} effect). This pattern has an amplitude of around 2 per cent in blue light, is not visible in red light, and is seen most clearly in the ratio of the blue to red master flats, where wavelength independent pixel-to-pixel sensitivity variations cancel out. A portion of one such ratio image is plotted in Fig.\,\ref{fig-lab} (left panel). 

The same light source was used without the lens or filter to flood the CCD with light in order to measure the illumination function of the camera shutter. We used a broad range of exposure times and the formulation of \citet{Zissell00} to determine the difference in exposure times with position on the CCD. An example shutter map is shown in Fig.\,\ref{fig-lab} (right panel).

High quality master bias and dark frames were also measured in the laboratory, the gain and linearity were measured, and hot pixels were mapped so that affected photometric points could be flagged. 

\subsection{Telescope mounts}
\label{sec-mounts}
The NGTS telescopes are each mounted on an equatorial fork mount made by Optical Mechanics Inc.,\footnote{\url{http://www.opticalmechanics.com}} allowing them to be independently pointed and guided. The mounts are arranged in two rows of six telescopes running side by side along the East-West direction (see Fig.\,\ref{fig-tels}). The inter-telescope spacing was chosen such that no telescope can intercept the field of view of any other telescope for elevations above $30^\circ$. 

The mounts are made from anodised aluminium and are fitted with a custom declination axis ring that interfaces with a matching ring surrounding the telescope tube.  The two axes are fitted with zero-backlash friction drives and their orientation is sensed with optical encoders.
The axes are operated as a closed loop servo-actuated system in order to optimise the response to wind and other environmental noise. The specification for the blind pointing accuracy of the mounts is 15\,arcsec, with relative pointing to better than 0.5\,arcsec over a distance of $1.5^\circ$. The maximum slew velocity is in excess of $10^\circ$ per second.

Each telescope is polar aligned using the drift method and by making fine adjustments to the altitude and azimuth of the telescope baseplate.  A pair of micrometers is used to enable repeatable adjustments at the $10\mu$m level. Precise polar alignment is important in order minimise the motion of stars through the night due to field rotation, which cannot be corrected by autoguiding. 
A telescope pointing model is generated using a grid of 900 pointings, spaced evenly in altitude and azimuth, enabling pointing accuracy of $\leq2$ pixels over the observable sky. The alignment of the telescope is quantified by analysing the pointing model data with TPoint\footnote{\url{http://www.tpointsw.uk}} and our design requirement is to maintain alignment to 
within 30\,arcsec of the celestial pole in order to keep field rotation below 1\,pixel at the edge of the field. In practice 
we align the mounts to 
$\sim 5$\,arcsec
from the celestial pole. As Chile is seismically active, we plan periodic checks of the alignment of each mount.

The low level mount control uses Clear Sky Institute Motion Controller (CSIMC) cards on the right ascension and declination axes. CSIMC cards are usually operated with the Talon Observatory Control System,
which is capable of controlling a complete observatory, but is not designed for a system with multiple telescopes in one building. We have therefore replaced large sections of Talon with custom software to provide global control of NGTS. A thin layer of Talon remains, essentially as an Application Programming Interface (API)
between our custom software and the CSIMC cards. We have also made our own modifications to the CSIMC firmware in order to enable continuous tracking and guiding on our fields for long periods. 

The mounts were supplied with limit switches that inform the CSIMC cards and hence our control software when an axis goes out of safe limits, but we have also fitted our own fail-safe system that cuts the power to a mount if either axis goes beyond hard limits. This security system can only be reset manually.

\subsection{Telescope enclosure and infrastructure}
\label{sec-facility}
The selected site for the observatory is 900\,m downhill from the VISTA telescope at an altitude of 2440\,m. 
A pre-existing dirt road links the NGTS facility to the rest of the ESO Paranal observatory. The NGTS enclosure sits on a concrete pad measuring $15\times15$\,m. The twelve telescope piers are cast into the inner section of the pad and are isolated from the surrounding concrete in order to minimise transmission of vibration. The telescope enclosure measures $15\times7$\,m and was supplied by GR~PRO\footnote{\url{http://www.grpro.co.uk}}. It consists of a metallic support structure that is surrounded by a fibreglass composite material. The roof is split into two halves that move apart along the North-South direction (see Fig.\,\ref{fig-ngts}. The roof panels are driven by a chain mechanism, which can be operated under battery power in the event of a power cut, and the roof can also be closed manually. 
The facility has a further two buildings; a converted shipping container control building that contains two server racks and office space; and a smaller transformer building that connects NGTS to the power grid at Paranal.

Overarching control of the observatory is by our own software control system, Sentinel, which monitors the global status of the facility (weather, network, mains power etc) and provides the final go/no-go decision to open the roof and begin observations. Sentinel continues to monitor global status during the night and automatically ceases observations and closes the roof when necessary. The roof is controlled via a Programmable Logic Controller (PLC) made by Beckhoff that communicates with Sentinel via the modbus TCP protocol. The PLC automatically closes the roof if communication with Sentinel is lost. 

The twelve individual telescopes are controlled by separate instances of our own telescope control system, Paladin, which is responsible for the control of the camera, focuser and mount. 
When allowed by Sentinel, the Paladins collect observing jobs from the operations database (described in Sect.\,\ref{sec-dms}) and act independently of each other. 
Sentinel and each of the twelve Paladins run on rack-mounted Linux servers situated in the control building. 

NGTS is equipped with a variety of sensors to ensure safe robotic operation. These include redundant mechanical and proximity sensors that detect the roof status.  A Vaisala WXT520 weather station that monitors temperature, pressure, wind, humidity and rain is installed on the roof of the control building, along with an AAG Cloudwatcher sky temperature probe. The Cloudwatcher also contains a light sensor and an additional rain sensor. As the detection of rain is always post-facto - and the NGTS roof takes approximately two minutes to close - we have chosen to install multiple sensors around the facility to permit the earliest detection of the first rain drops. This includes an additional bank of sixteen $5\times3$\,cm rain sensors on the roof of the control building that are connected to a Raspberry Pi, and a further rain sensor connected directly to the PLC inside the telescope enclosure (bringing the total to 19). 
A Dylos dust sensor is installed in a weatherproof box outside on the East wall of the telescope enclosure. 

A monochromatic Alcor OMEA all-sky camera is installed on the control building roof and permits the early detection of incoming clouds. We have also installed eight AXIS network cameras to monitor the facility, including three low-light level cameras that allow us to monitor the status of the telescopes and the enclosure roof even in dark sky conditions. Network microphones have also been installed to provide additional remote monitoring of the roof mechanism. 

Equipment in the telescope enclosure is connected to servers in the control building via a multicore fibre bundle (a distance of $\geq 20$\,m). A pair of fibres in the bundle also provides the network connection to ancillary devices in the telescope enclosure (webcams, network addressable power distribution units, PLC etc). The fibre connection is converted to USB 2.0 at each end using a pair of Icron Ranger USB-to-fibre converters. 

\subsection{Data management system}
\label{sec-dms}
NGTS employs a database driven system for managing all aspects of observatory control and data management. This centralises observatory operations and data analysis, allowing the efficient sharing of information between different stages of data collection, reduction and analysis (described in Sects.\,\ref{sec-ops},\,\ref{sec-reduction}\,\&\,\ref{sec-analysis}). There are 4 main MySQL databases, described below, one for each of operations, data tracking, data reduction and candidate tracking. 

Information required for observation scheduling, meta data such as the current time, pointing, focus, action type and autoguiding statistics, along with environmental data such as weather and Sun/Moon positions are stored in a series of tables in the operations database at Paranal. A subset of this information forms the FITS image headers.

The combined 12 telescopes of NGTS generate an average of 200\,GB of images per night, which compresses by around a factor of two with the 
\texttt{bzip} algroithm. 
Due to limited network bandwidth the data is transferred to the University of Warwick each fortnight via removable 2 TB hard discs. The data are ingested into the NGTS cluster and also backed up to larger 6 TB discs for safety. The 2TB discs are then reformatted and returned to Chile for reuse. A database driven tracking system spanning Paranal and the University of Warwick, ensures safe transfer of compressed FITS images from Chile to the archive in the UK. Only once an image is confirmed to exist in the UK archive, is it flagged for removal at Paranal.

Data products, such as raw photometry and image statistics from the data reduction pipeline (described in Sect.\,\ref{sec-reduction}) and detrended photometry (Sect.\,\ref{sec-analysis}), are  stored in the pipeline database at the University of Warwick. A data quality assessment web page sits on top of the pipeline database, allowing for checks of the data reduction pipeline output.  

The candidate database houses the measured properties of exoplanet candidates, external catalogues (for cross referencing purposes) and candidate summary statistics. The information on each candidate is displayed on a series of web pages (named Opis) where members of the consortium regularly convene to vet potential exoplanet candidates (internally known as \emph{eyeballing}). 

The two sites (Paranal \& University of Warwick) are synchronised across the network using SymmetricDS.  In the case of a network outage, SymmetricDS gathers all changes to the databases at each location and automatically syncs the system when the network connection returns.  

\begin{figure*}	\includegraphics{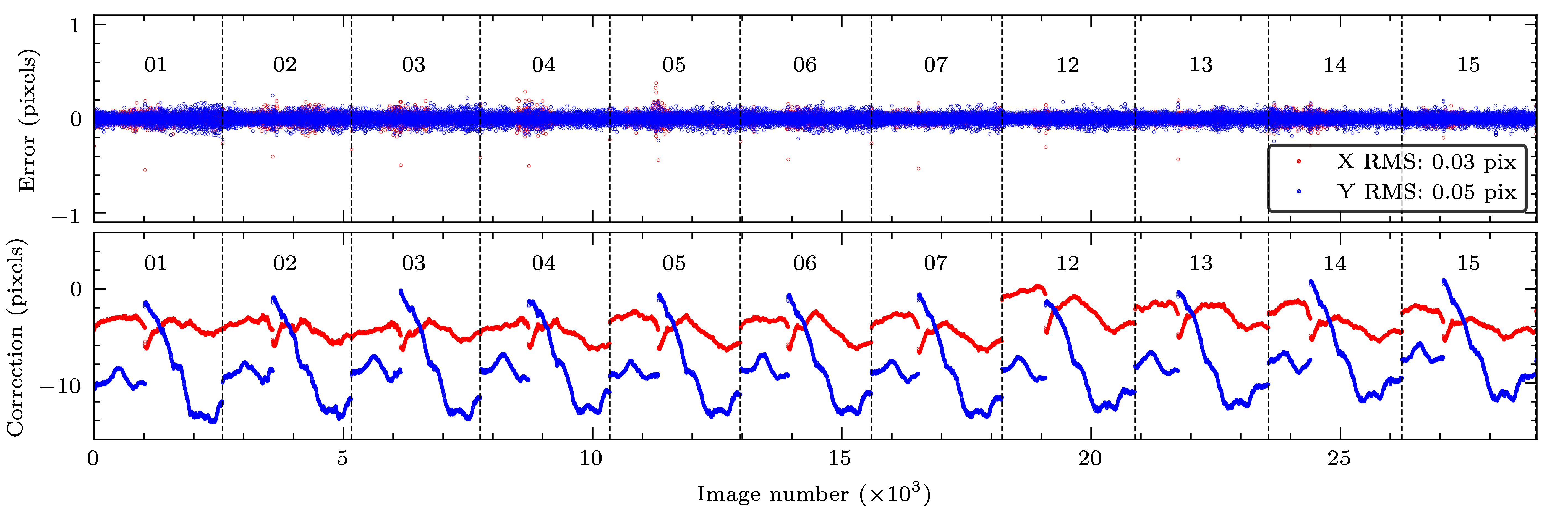}
    \caption{Top panel: Frame-to-frame autoguiding residuals for one telescope unit from 11 nights in March 2016. The RMS guiding error is $0.04$ pixels on average. Bottom panel: The cumulative guide correction applied to the telescope over the same 11 nights. 
The telescope mounts require absolute corrections, hence the cumulative error. 
The numbers along the top of each panel denote the day of the month in March 2016.}
\label{fig-guiding}
\end{figure*}

\section{NGTS Operations and survey}
\label{sec-ops}
The NGTS facility operates robotically, with no human intervention necessary, although we do require a human go/no-go decision each night as an additional safety measure. The roof opens one hour before sunset, allowing for equipment to settle to ambient temperature, and a sequence of approximately one hundred flat-field images are taken while the Sun is between altitudes of --4.5$^\circ$ and --8.5$^\circ$ with the telescopes pointing at an altitude of 75$^\circ$ at the anti-Solar azimuth in order to minimise brightness gradients \citep{Chromey96}. Flat fields are followed by a focus run to monitor the optimal focus offset for each camera, and we find the focus to be quite stable night-to-night, with adjustments needed only occasionally.  Science operations are carried out while the Sun is below an altitude of $-15^\circ$, and are followed immediately by a second focus run. A second set of flat field images are taken in morning twilight, after which the roof is closed and a sequence of dark frames and biases are taken while the ambient light level is low. 

During the night each of the 12 telescopes operates in either survey or follow-up mode.  In survey mode the telescope observes a sequence of pre-assigned survey fields, with each field followed continuously as long as it has the highest altitude. For our baseline survey we aim to space fields such that one field rises above $30^\circ$ elevation as the previous field sets below $30^\circ$. Thus each telescope typically observes two fields per night. Fields are followed with the same telescope every night that they are visible, providing the maximum coverage possible over a single observing season. This results typically in around 500\,h coverage spread over 250\,nights. 
Fields that pass within $25^\circ$ of the Moon on a given night are replaced with a back-up field.

In follow-up mode the telescope targets a particular star, which is placed at the centre of the field of view to minimise movement due to differential atmospheric refraction. For both modes the default is to observe in focus and with exposure times of 10\,s, but these choices can be manually configured. 

\subsection{Survey field selection}
Survey fields for each telescope are selected manually from a mesh of 5\,307 field centres that efficiently cover the entire sky (overlaps of 3 per cent on average). Fields are selected against criteria that take into account the density of stars, the proportion of dwarf stars, the ecliptic latitude and the proximity of very bright and extended objects. 

To aid this selection we have carried out our own sky survey with NGTS covering all 3\,540 southern fields visible to our telescopes. We use this survey to assess the number of unblended target stars in each field, using an empirical measure of the dilution of light from each star by its neighbours. We find our survey images are also useful in assessing the impact of scattered light from bright stars that can be outside the field of view. 

In addition to the number of stars that appear unblended in our NGTS images, we consider the expected rates of false positive transit detections due to faint background objects \citep{Guenther17}. 
We also cross-match our source lists with the PPMXL proper motion survey \citep{Roeser10} and 2MASS photometry \citep{Skurtskie06}, allowing us to use reduced proper motion to estimate the proportion of dwarf and giant stars in each field \citep{Collier-cameron07}. Typically we select fields with $\leq 15\,000$ stars brighter than an $I$ band magnitude of 16, of which $\geq 70$ per cent are dwarf stars. These fields are usually more than $20^{\circ}$ from the Galactic plane. We also tend to avoid fields within $30^{\circ}$ of the ecliptic plane, because they are adversely affected by the proximity of the Moon for about 3 nights per month. 

\begin{figure*}
\includegraphics{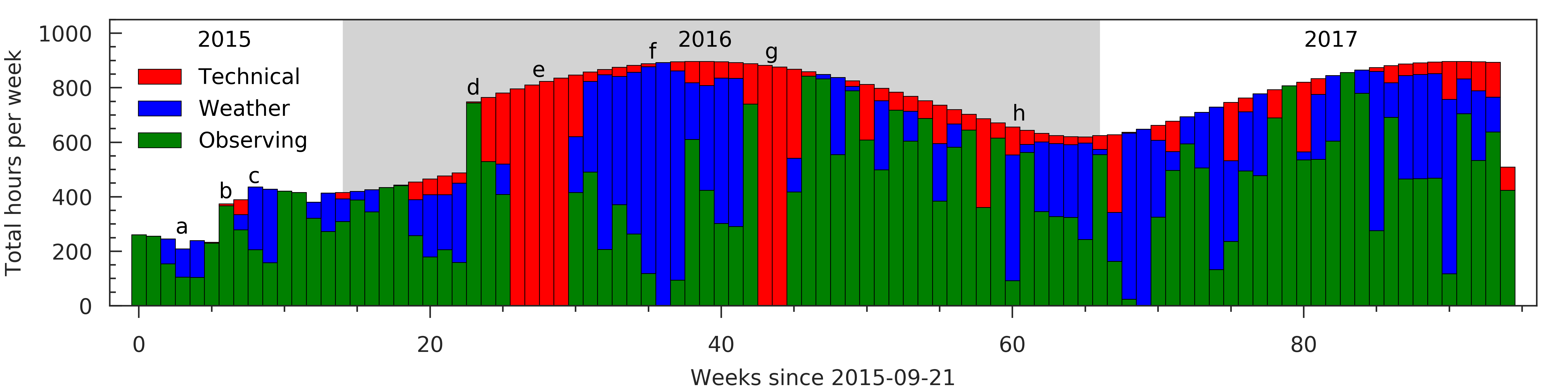}
\caption{NGTS observing statistics since the first survey commissioning data were taken with 4 telescope units on 2015 September 21. The time available on each night was calculated as the number of hours between astronomical twilights multiplied by the number of installed telescopes. The colours green, blue and red denote time spent observing, time lost to bad weather and time lost due to technical issues, respectively.   Several key dates are marked on the plot as follows: a) survey commissioning observations began with 4 telescopes; b) 3 more telescope units installed, c) 1 more unit installed, bringing total to 8; d) installation completed for the final 4 telescope units; e) NGTS was invaded by rodents and cabling was destroyed resulting in 1 month of technical downtime; this downtime also marks the end of commissioning observations and the beginning of full survey operations; f) Paranal suffered particularly poor weather during May, June and July due to an El Ni{\~n}o event; g) NGTS suffered a further 2 weeks of technical downtime due to a fault with the enclosure roof; h) individual telescope units were off-sky for extended periods 
due to ongoing camera shutter lifetime issues. 
}
\label{fig-stats}
\end{figure*}

\subsection{Autoguiding}
\label{sec-guiding}
A key component of our strategy to achieve high photometric precision is to 
minimise the movement of stars on the CCD detector (Sect.\,\ref{sec-goals}). 
We do this by employing a closed-loop guiding system that operates on the science images in real time. Our algorithm is an updated version of the DONUTS autoguiding system described by \citet{McCormac13}. This uses a master reference image to re-acquire a given field to the same sub-pixel position as on previous nights, and monitors that position using a series of 1D cross correlations between the science images and the reference images. Guiding corrections are passed through a proportional-integral-derivative control loop to smooth the corrections and allow for the response of the telescope drive. A necessary modification to DONUTS was to detect and mask out the lasers used as guide stars at the VLT, which otherwise dominate the cross correlations and cause the autoguider to lose lock. We also implemented a filter to ignore spurious offsets caused by aeroplanes passing through the field of view. 

Our typical guiding performance is illustrated in Fig.\,\ref{fig-guiding} where we show the guiding residuals for one telescope unit over a period of eleven nights. We achieve excellent sub-pixel tracking of the field with an average RMS in the frame-to-frame offsets of only $0.04$ pixels (upper panel). In the lower panel of Fig.\,\ref{fig-guiding} we show the guiding corrections applied by the autoguiding algorithm, which would be the distance moved by the stars without guiding. 
The structure in this cumulative correction stems primarily from telescope tracking inaccuracies and tube flexure, with possibly some residual polar misalignment.

As outlined in Sect.\,\ref{sec-goals}, differential refraction across our wide field of view still causes star positions to shift slightly, despite this superb autoguiding performance. There is no shift at the centre of the field of view, where we place the target stars in follow-up mode, but the shift increases to a maximum of 0.75 pixel 
between elevations of 30$^\circ$ and 90$^\circ$ at the edges of the field of view.
The shift acts along the parallactic angle, which rotates with respect to our field of view during the night for most fields. 

A Python implementation of our upgraded autoguiding algorithm is available via PyPi\footnotemark\footnotetext{\url{https://pypi.python.org/pypi/donuts}} and can be found on GitHub\footnotemark\footnotetext{\url{https://github.com/jmccormac01/Donuts}}.

\subsection{Real-time monitoring}
The autoguider statistics along with many other indicators of instrumental health and data quality are written in real time to the operations database of the data management system (Sect.\,\ref{sec-dms}) and can be monitored remotely via a web interface. This includes outputs from the many sensors described in Sect.\,\ref{sec-facility}, including weather sensors and the status of the enclosure roof. Our Paladin telescope control systems also carry out real time monitoring of science images, including sky background level and structure, stellar image size across the field of view, and detection and flagging of the VLT laser guide stars. 

We also carry out a basic photometric data reduction in real time in order to measure the transparency of the atmosphere along the line of site of each telescope. Aperture photometry is measured for each star and is compared to the fluxes measured in the autoguider reference image (obtained in good conditions).  A percentage difference in the atmospheric throughput is recorded and this value is used to determine the photometric quality of a given night. When highly non-photometric conditions are recorded the facility is sometimes closed manually as a precaution against unexpected rain. 

\subsection{Observing statistics}
As part of our monitoring of the NGTS facility we track any time lost due to weather and technical issues and produce weekly statistics. Figure\,\ref{fig-stats} shows the operations statistics since survey commissioning observations began with four telescope units on 2015 September 21. The remaining eight telescope units were installed over three further missions in 2015 and Feb 2016 (marked b, c and d on Fig.\,\ref{fig-stats}). Unfortunately considerable time was lost during summer 2016 due to unusually poor weather at Paranal associated with an El Ni{\~n}o event (marked blue in Fig.\,\ref{fig-stats} and labelled f). We also suffered two periods of technical downtime due a rodent infestation and a fault with the enclosure roof mechanism (marked red in Fig.\,\ref{fig-stats} and labelled e and g respectively). Other time lost to technical issues is primarily due to failures of shutters on individual cameras. During week 71 we modified the camera shutter drivers with the goal of improving the shutter lifetime.

The extended downtime due to rodents in March and April 2016 (labelled e) also marks the end of commissioning observations and the beginning of full survey operations.

\section{Data reduction}
\label{sec-reduction}
NGTS data are reduced using a custom built pipeline that is called within the data management system running at the University of Warwick (Sect.\,\ref{sec-dms}). The pipeline is modular, with each task being called separately as needed. For each observing field we begin by generating a catalogue of target stars (Sect.\,\ref{sec-catalogue}). Each night of science images for each field is then bias subtracted and flat fielded (Sect.\,\ref{sec-cal}), astrometric solutions found (Sect.\,\ref{sec-astrometry}) and photometric measurements made (Sect.\,\ref{sec-phot}). A set of light curves for each field-season are then assembled and made available for downstream detrending and transit searches (Sect.\,\ref{sec-analysis}).

Breaking the pipeline into smaller modules in this way helps ensure efficient use of computing resources, allowing us to cope with the relatively high data rates and to plan for reprocessing of data with improved algorithms. The {\sc SGE}\footnote{Sun Grid Engine, now Oracle Grid Engine} scheduling system is used to interleave jobs at various stages of the pipeline across our compute cluster with minimal deadlocks. Each pipeline module is also internally parallelised in order to further improve processing efficiency. 

\subsection{Catalogue generation}
\label{sec-catalogue}
For each NGTS survey field we carry out our own source detection and generate our own catalogue of target stars. This avoids the risk of misplaced source apertures due to proper motion,
which would disproportionately affect M-dwarfs for which the smallest exoplanets should be detectable. Using our own catalogue does mean that some known blended stars are not resolved in our source catalogues, but only where the light curves of the blended stars cannot be fully separated. 

As the NGTS images are undersampled, the source detection for each field is carried out on a stacked master image that is made from a sequence of images with deliberate dithering between exposures. This improves the astrometry by better sampling the stellar profiles. One hundred images are taken while the field is at low airmass using offsets of around 30\,arcsec (6\,pixel) and 10\,s exposures. 
The images are supersampled, aligned using our autoguider algorithm (Section~\ref{sec-guiding}) and then averaged to produce a deep and high resolution master image. 
The stacked image is then solved astrometrically (Section~\ref{sec-astrometry}) and the source detect performed using {\sc imcore} from the \casu{} software suite\footnote{\url{http://casu.ast.cam.ac.uk/surveys-projects/software-release}}~\citep{Irwin04}.

Sources are detected in the dithered stack down to I band magnitudes of around 19, but we limit our standard source catalogues to $I<16$, which is close to the detection limit in a single 10\,s exposure. Fainter objects can be added manually to the target list as required. 

Each detected source is cross-matched with a number of other catalogues including the
AAVSO Photometric All-Sky Survey \citep[APASS;][]{Henden14}, \gaia\ \citep[][]{Gaia2016}, 2MASS \citep[][]{Skurtskie06} UCAC4 \citep[][]{Zacharias13}, ALLWISE \citep[][]{Cutri14}, RAVE \citep[][]{Kunder17} and GALEX \citep[][]{Martin05}.
During cross matching with APASS, \gaia\ and 2MASS we apply empirically defined limits on colour and separation to avoid spurious matchings. The matching with ALLWISE and RAVE is carried out via the 2MASS ID of each source. The APASS matches are used to compute an approximate $I$-band zero point for each field in order to set the faint limit of the target list. We use the \gaia\ cross match to determine whether each NGTS source is a single object or a blend that is unresolved in NGTS images. 
For high proper motion stars we currently use UCAC4 data to improve cross matching between catalogues, however we plan to use \gaia\ proper motions once these are available.

\subsubsection{Stellar type estimation}
\label{sec-sed}
As part of the generation of the target catalogue for each survey field we perform a 
preliminary spectral classification of each star. 
The classification is used in the vetting of exoplanets candidates (Sect.\,\ref{sec-vetting}) and is potentially useful for a wide range of variable star studies. 

For each star we determine the most likely spectral type, luminosity class and interstellar reddening by fitting the spectral energy distribution (SED) formed from the full set of available magnitudes (Sect.\,\ref{sec-catalogue}). The fit is performed by finding the minimum $\chi^{2}$ between the observed photometry and a grid of synthetic magnitudes for main sequence and giant stars. The synthetic photometry was derived by convolving the filter profiles with the stellar spectra library by \citet{Pickles98}, which we reddened using the standard $R_{V} = 3.1$ law by \citet{Fitzpatrick99}. For each NGTS source, we limited the grid of reddened synthetic photometry to the maximum line-of-sight asymptotic reddening by
\citet{sfd98}.
In our SED fitting procedure, we also take into account the dwarf/giant probability for each source from its position in a reduced proper motion diagram \citep{Collier-cameron07} and estimate photometric parallaxes using the absolute magnitude scale presented in \citet{Gray09}. Spectral type, luminosity class, reddening and distance, are all included in the source catalogue. 

This method 
will be refined once \gaia\ parallaxes are available for our target stars.

\subsection{Image reduction and calibration}
\label{sec-cal}
Science images are bias-subtracted and flat-field corrected using standard procedures.  Bias and dark frames are acquired at dawn after the enclosure roof has closed, and twilight flat-field frames are acquired at both dawn and dusk (Sect.\,\ref{sec-ops}).  Each image is first overscan subtracted using columns robust to bleeding, as determined by the lab characterisation (Sect.\,\ref{sec-lab}).  Bias residual frames are then mean combined to produce 
master bias frames.  Dark frames are not subtracted during the reduction process as the dark current is negligible, but master dark frames are monitored.  Twilight flat-field frames are sigma-clipped to remove stars and mean combined.
Shutter maps are obtained following the method from \citet{Surma1993} and are monitored for indications of shutter failure.  A full observing season's worth of bias and flat-field action master frames, with outlier rejection, are used to construct the best 
overall calibration master frames for science images. The quality and variation of flat-field frames over time is monitored, and new master flats are constructed after hardware maintenance (when a camera shutter has been replaced for example).

\subsection{Astrometry} 
\label{sec-astrometry}
For each NGTS science image we find a full World Coordinate System (WCS) astrometric solution, which we store in the standard FITS keywords \citep{Greisen02}.
This enables precise placement of photometric apertures for each target star.
An astrometric solution is needed for each image despite the precise autoguiding of the NGTS telescopes (Sect.\,\ref{sec-guiding}) in order to account for field stretching due to differential atmospheric refraction (Sect.\,\ref{sec-goals}) and any field rotation due to imperfect polar alignment (Sect.\,\ref{sec-mounts}). 

The NGTS telescopes have non-linear radial distortion, and so we chose to use the zenithal polynomial (ZPN) projection \citep{Calabretta02}. We found it necessary to use a 7th order polynomial, with the distortion described by the 3rd, 5th and 7th terms (PV2\_3, PV2\_5 and PV2\_7 WCS keywords). 
The distortion is stable with time, so we measure it once for each telescope and keep the distortion model fixed when solving individual images. The distortion model is only revisited after hardware maintenance (e.g.\ refitting of a camera after a shutter replacement). 

The radial distortion is measured using our own code that employs a Markov chain Monte Carlo (MCMC) method\footnote{{\sc emcee}: \url{http://dan.iel.fm/emcee/current/}} to find the polynomial coefficients and the pixel coordinates of the centre of the distortion. Individual images are then solved for translation, rotation, skews and scales using the {\sc wcsfit} program from the \casu{} software suite \citep[][with the results stored in the CDi\_j WCS FITS keywords]{Irwin04}. Both programs use the 2MASS catalogue for the reference astrometry. An initial approximate solution for each field is found using astrometry.net \citep{Lang10}.

\subsection{Photometry}
\label{sec-phot}
Our photometric measurements are made using aperture photometry with the \casu{} \mbox{{\sc imcore\_list}} program \citep{Irwin04}. For each star in our input catalogue (Sect.\,\ref{sec-catalogue}) we define a soft-edged circular aperture with a radius of 3 pixels (15\,arcsec) and these are placed in pixel coordinates using our per-image astrometric solutions (Sect.\,\ref{sec-astrometry}).
The sky background for each pixel in the source aperture is estimated using bilinear interpolation of a grid of $64\times64$ pixel regions for which the sky level is determined using a k-sigma clipped median.

Although not routinely applied, the NGTS pipeline also allows for difference imaging before aperture photometry 
using a method based on the {\sc ISIS} code by \citet{Alard2000}. We found that for fields with typical crowding there was no clear advantage to image subtraction, as was expected for our under-sampled images, but this remains an option for more crowded fields. 
Due to our precise autoguiding (Sect.\,\ref{sec-guiding}) it is generally not necessary to register images before applying the image subtraction. 

\section{Data analysis and transit search}
\label{sec-analysis}
Once data for a given field have been reduced and photometric measurements made for each science image (Sect.\,\ref{sec-reduction}) we assemble a light curve for each target star, detrend for red noise sources 
(Sect.\,\ref{sec-detrend}) and search for exoplanet transits (Sect.\,\ref{sec-transits}). Detected signals are subjected to a number of vetting tests 
(Sect.\,\ref{sec-vetting}) before the best candidate exoplanets are followed up with further photometric and spectroscopic observations (Sect.\,\ref{sec-followup}).

\subsection{Light curve detrending}
\label{sec-detrend}
To detrend the photometric data from systematic signals, we use several detrending algorithms. To correct first order offsets, common to all light curves, a mean light curve is calculated and used as an artificial standard star for correcting all the stars.  This is the first step of our own implementation of the SysRem algorithm \citep{Tamuz05}, which is an updated version of that used by the WASP project \citet{Cameron06}. SysRem removes signals that are common to multiple stars, even where the amplitudes of the signals vary between stars.

Additionally we found systematic signals that correlate with  Moon phase and sidereal time, which have different shapes for different stars, are not completely removed by SysRem. The signals related to Moon phases are likely to reflect imperfect sky subtraction and/or low-level non-linearity of the detectors. Sidereal time is degenerate with airmass, as well as sub-pixel movements of stars due to differential atmospheric refection, and so systematics correlating with sidereal time might arise from differential extinction, imperfect flat fielding and/or sub-pixel sensitivity variations. 

To correct for such periodic systematics and to allow for removal of periodic stellar signals (which are not noise but might still  prevent us from detecting transit signals) we perform an analysis of variance to identify significant periodic signals. After verifying that the detected signal does not have a transit shape these signals are removed by calculating the floating mean in the phase domain (a detailed description can be found in Eigm\"uller et al. in preparation). In addition we have tried detrending with x and y pixel position with similar results. 

We found that correcting for periodic signals improves our transit detection efficiency by 10--30 per cent and decreased the number of false detections by 50 per cent (see Sect.\,\ref{sec:det_eff}.

\subsection{Transit detection}
\label{sec-transits}
\label{sec:orion}
After de-trending, the NGTS light-curves are searched for transit-like signatures using a Box-Least-Squares (BLS) algorithm. The code, called {\sc orion}, has been used for most of the transit detections of the WASP project and is described in more detail by West et al.\ (in preparation). It is based on the formulation of \citet{Cameron06} with a number of key enhancements that improve the sensitivity and speed of the transit search. Foremost amongst these is an extension to allow for the fitting of box profiles of multiple widths (from 1.5\,hr to 3.75\,hr in steps of 0.75\,hr) in order to better match the transit signatures of planets in inclined orbits. With an appropriate re-casting of the original formulation this 
was 
achieved with minimal loss in speed. {\sc orion} can combine data from multiple cameras, survey fields and observing seasons. It also incorporates the Trend Filtering Algorithm (TFA) de-trending from \citet{Kovacs05}. The code is parallelized using OpenMP, and scales well to high core-count.

We also plan to use the DST algorithm \citep[D\'etection Sp\'ecialis\'ee de Transits;][]{Cabrera12} 
which provides a better description of the transit shape with the same number of a free parameters as BLS. DST also allows a more flexible definition of the region in transit, which is useful for taking into account transit timing variations \citep[see also][]{Carter13}.
The experience of the \corot\ community was that applying independent transit detection algorithms to the same data maximised the number of transit detections and facilitated the identification of false positives \citep{Moutou05,Moutou07}.

\subsection{Planet candidate vetting}
\label{sec-vetting}
For vetting of candidates we aim to automate the procedure as much as possible to ensure repeatable outcomes and best possible performance.  We use an automated vetting algorithm named CANVAS (CANdidates Vetting, Analysis and Selection) which identifies the signals detected by {\sc Orion} (see Section \ref{sec:orion}) that are most consistent with a transiting planet signal.  
CANVAS first fits the \citet{Mandel02} transit model to each feature detected by {\sc Orion} using the {\sc BATMAN} 
code 
\citep{Kreidberg2015}. Combined with estimated stellar parameters from SED fitting (Sect.\,\ref{sec-sed}) this provides putative planet radii, impact parameters, orbital separations and stellar densities. 
CANVAS then
down-weights
detections with common periods (usually arising from systematics), detections with poor phase coverage during transit,
and detections from light curves with large amplitude variability (usually variable stars).  
The NGTS light curve is also used to check whether secondary transit events are visible, or if a difference between odd and even transit events can be spotted. Either would suggest the observed signal is caused by an eclipsing binary. Using the transit fitting and SED results, 
together with information from
the Besan\c con galaxy model \citep{robin03}, we also 
assess
the plausibility of the planet hypothesis using the stellar density \citep{Tingley2011}.

In addition to the CANVAS diagnostics, we assess the significance of the transit feature by sliding the transit model through phase space and computing the likelihood at each step. This method is also adept at identifying eclipsing binaries through their secondary eclipses. We 
model and compute the likelihood of individual transits to check that the signal increases with additional transits in the manner expected for a genuine occultation as opposed to correlated noise.

For the first time in a ground-based transit survey, NGTS also employs automated centroid vetting 
\citep{Guenther17a}. This is important because background eclipsing binaries blended in the photometric aperture can mimic planetary transits, and our modelling has shown that such signals are four times more frequent than planet transits for NGTS, making them very costly in follow-up time \citep{Guenther17}. The centroiding technique detects the small shift in flux centroid towards the target star when off-centre flux is lost during the eclipse of a blended binary. We reach a precision of $< 1$~milli-pixel on average over an entire field, and as low as $0.25$~milli-pixel for specific targets. We estimate that this enables the identification of more than $50$~per-cent of background eclipsing binaries without requiring follow-up observations. Additionally, the centroiding technique provides the undiluted depth of any transit signal, preventing misclassification of planet candidates. Our full method is described by \citet{Guenther17a}.

We are also developing a machine-learning based autovetter to further automate the candidate vetting process. This will incorporate all of the above information to provide ranked lists of candidates, prioritising those most likely to represent true transiting planets in a systematic and repeatable fashion. While not yet finalised, proven algorithms such as Random Forests \citep{McCauliff15a} and self-organising-maps \citep{Armstrong17a} are being explored.

The results of the various vetting procedures are ingested into the candidates database of the NGTS data management system and can be interactively interrogated using our Opis web interface (Sect.\,\ref{sec-dms}). The most promising candidates are flagged for follow up observations (Sect.\,\ref{sec-followup}). 

\begin{figure}
	\includegraphics[width=8cm]{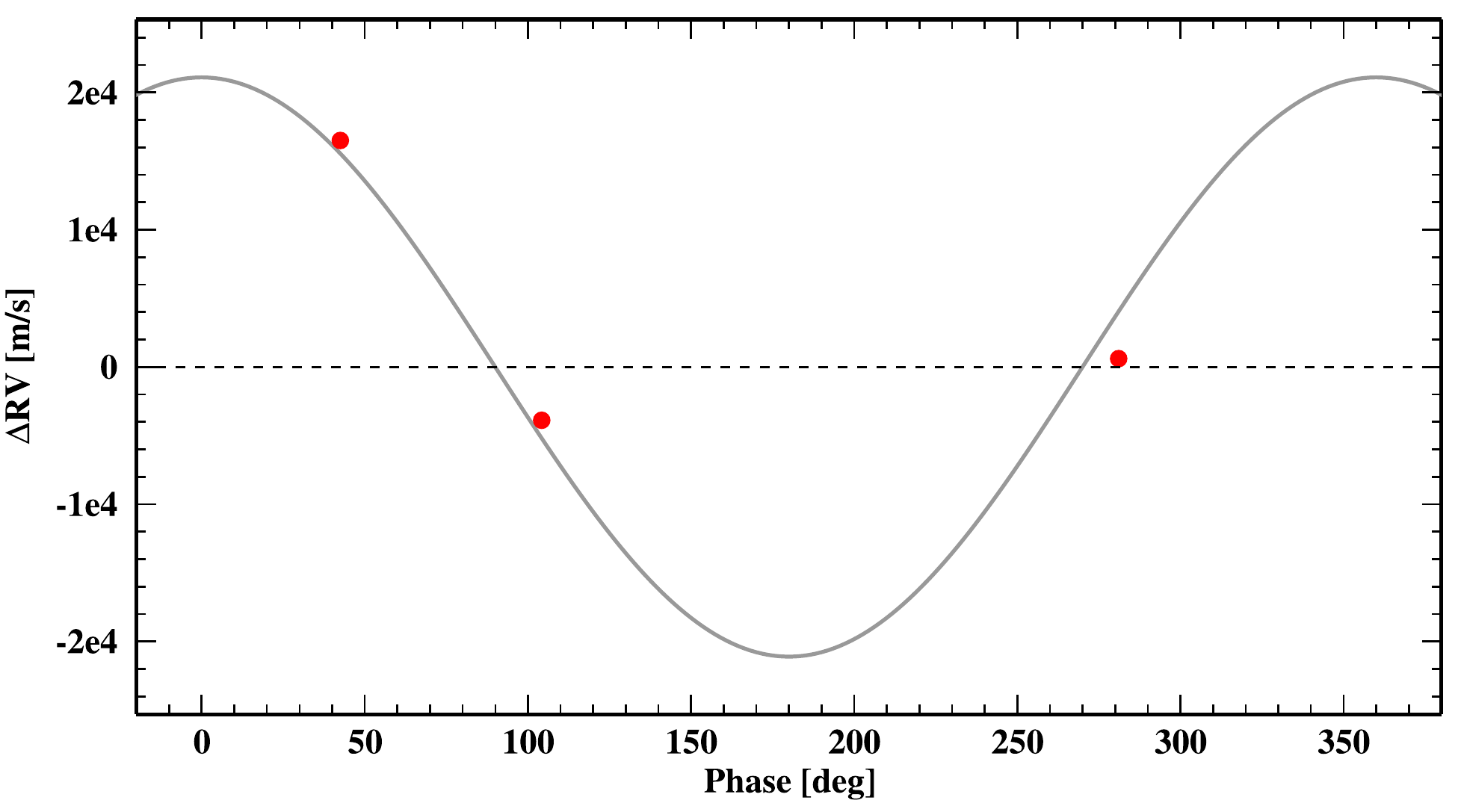}
    \includegraphics[width=8cm]{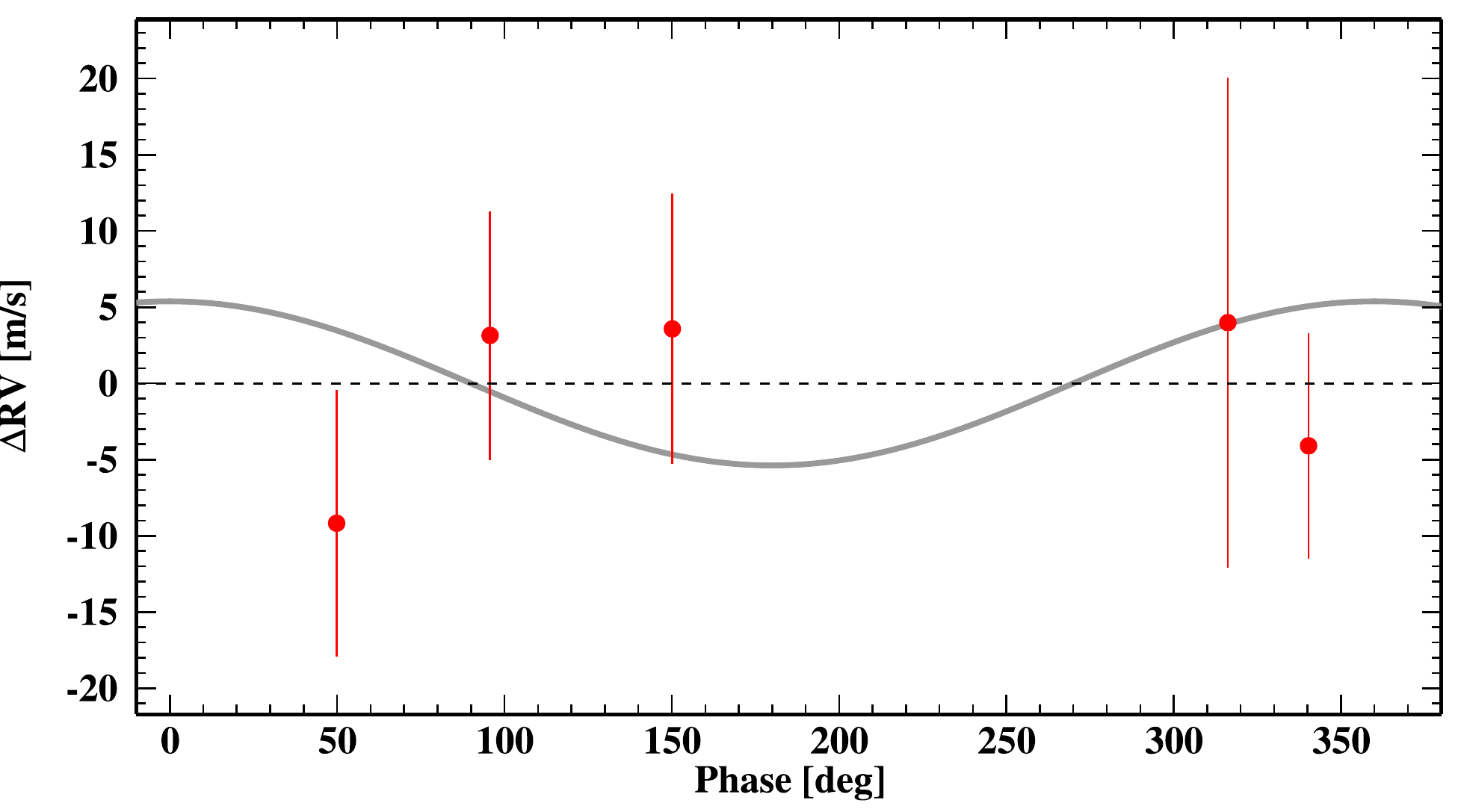}
    \caption{CORALIE radial velocity measurements. \textit{Top} NGTS candidate NG0531-0826-35017 phase-folded using the photometric period ($P=5.70232$\,d) and phase ($T_{c}=2457291.7583$). \textit{Bottom:}  NGTS candidate NG1947-4200-11647 phase-folded using the photometric period ($P=1.29297$\,d) and phase ($T_{c}=2457289.537789$).  For both plots the red circles are individual CORALIE measurements (uncertainties smaller than point size in top plot) and the solid line is a best fit Keplerian orbit with $e=0$ and period and phase fixed at stated values.}
    \label{fig:coraliervs}
\end{figure}

\section{Follow up observations}
\label{sec-followup}
Transit candidates that survive the vetting described in Sect.\,\ref{sec-vetting} are passed to CORALIE for spectroscopic vetting (Sect.\,\ref{sec:CORALIE}) and then for radial velocity follow up with FEROS and HARPS (Sects.\,\ref{sec:FEROS}\,\&\,\ref{sec:HARPS}). System parameters are determined from joint fits to light curves and radial velocity measurements (Sect.\,\ref{sec-params}).

\subsection{Candidate vetting with CORALIE}
\label{sec:CORALIE}
We spectroscopically vet candidates using the CORALIE spectrograph \citep{Queloz00} on the 1.2\,m Euler Telescope at La Silla Observatory, Chile.  CORALIE is a high-resolution ($R \sim 50,000$) fibre-fed echelle spectrograph designed for high precision radial velocity measurements.  For bright stars, the long term radial velocity precision of CORALIE is $<6$\,\ms \citep{Marmier13}.  For NGTS candidates, with a mean magnitude of $V=13.5$, the radial velocity precision is photon limited, and we typical achieve 20-30\,\ms~with a 30-45\,min exposure time.  CORALIE has a long history of being used to confirm transiting exoplanets, most notably for the WASP survey \citep{Pollacco06}.  The primary differences in terms of monitoring NGTS targets is that they are 
typically fainter than WASP candidates, and the expected planet masses can be considerably lower.
The combination of these factors means that for NGTS candidates, CORALIE is mainly used to vet candidates rather than provide confirmation and mass determination - although this is possible for some hot Jupiters discovered by NGTS.

Observations of an NGTS candidate begins with a single spectrum, preferably acquired at the expected maximum or minimum radial velocity phase ($\mathrm{phase}=0.25$ or 0.75).  The guider camera image is inspected for evidence of a visual binary which may not have been apparent in the NGTS or archival imaging.  The data are reduced with the standard CORALIE data reduction pipeline, and we inspect the resulting cross-correlation function (CCF) for evidence of two peaks indicative of a binary star system.  We also check that the CCF is not broadened (due to rapid rotation of the star) which would make precise radial velocity measurements difficult.  If the CCF is single-peaked and not broadened, we acquire further epochs spanning the orbital phases.  We fit the resulting multi-epoch radial velocity measurements with a zero-eccentricity Keplerian model, fixing the period and phase from the NGTS photometric discovery data.  This provides a mass estimation for the companion object, or a mass limit where no variation is seen above the level of the measurement uncertainties.  Data are archived and analysed using the DACE platform. \footnote{\url{https://dace.unige.ch}}  As examples, CORALIE radial velocity measurements for candidates NG0531-0826-35017 and NG1947-4200-11647 are shown in Fig.~\ref{fig:coraliervs}.  NG0531-0826-35017 displays a high amplitude ($K=21~\kms$) in-phase variation indicative of an eclipsing binary.  NG1947-4200-11647 shows no variations $>5$\,\ms\ ruling out a high-mass planetary companion and warranting higher precision monitoring with FEROS and/or HARPS (see Sects.\,\ref{sec:FEROS}\,\&\,\ref{sec:HARPS}).

\subsection{Radial velocity monitoring with FEROS}
\label{sec:FEROS}
The Fibre-fed Extended Range Optical Spectrograph (FEROS; \citealp{Kaufer99}) 
is a high-resolution ($R \sim 48,000$) echelle spectrograph that maintains a very high throughput of light ($\sim$20\% total efficiency), covering almost the entire optical spectral range (3700 -- 9000\AA).  FEROS is mounted on the MPIA 2.2\,m at La Silla Observatory, Chile.  Calibration and reduction of the observed data with this instrument uses the pipeline procedure CERES \citep{Brahm17}, where typical echelle spectral calibration routines are performed, such as debiasing, flat-fielding using the illumination from a halogen gas lamp, scattered-light removal, and wavelength calibration. The pipeline also measures radial velocities and bisector spans, and \citet{Brahm17} have shown FEROS to have a long-term stability at the $\approx 8\,\rm m\,s^{-1}$ level for bright dwarfs. Work measuring precision radial velocities of giant stars has shown FEROS to be stable at a similar level \citep{Soto15,Jones16}.

The increased telescope aperture compared to CORALIE means that FEROS can reach a higher radial velocity precision at the brightness of typical NGTS target stars, therefore NGTS candidates vetted with CORALIE may be passed to FEROS for further vetting or mass and orbit characterisation.

\subsection{Radial velocity follow up with HARPS}
\label{sec:HARPS}
\begin{figure}
    \includegraphics[width=8cm]{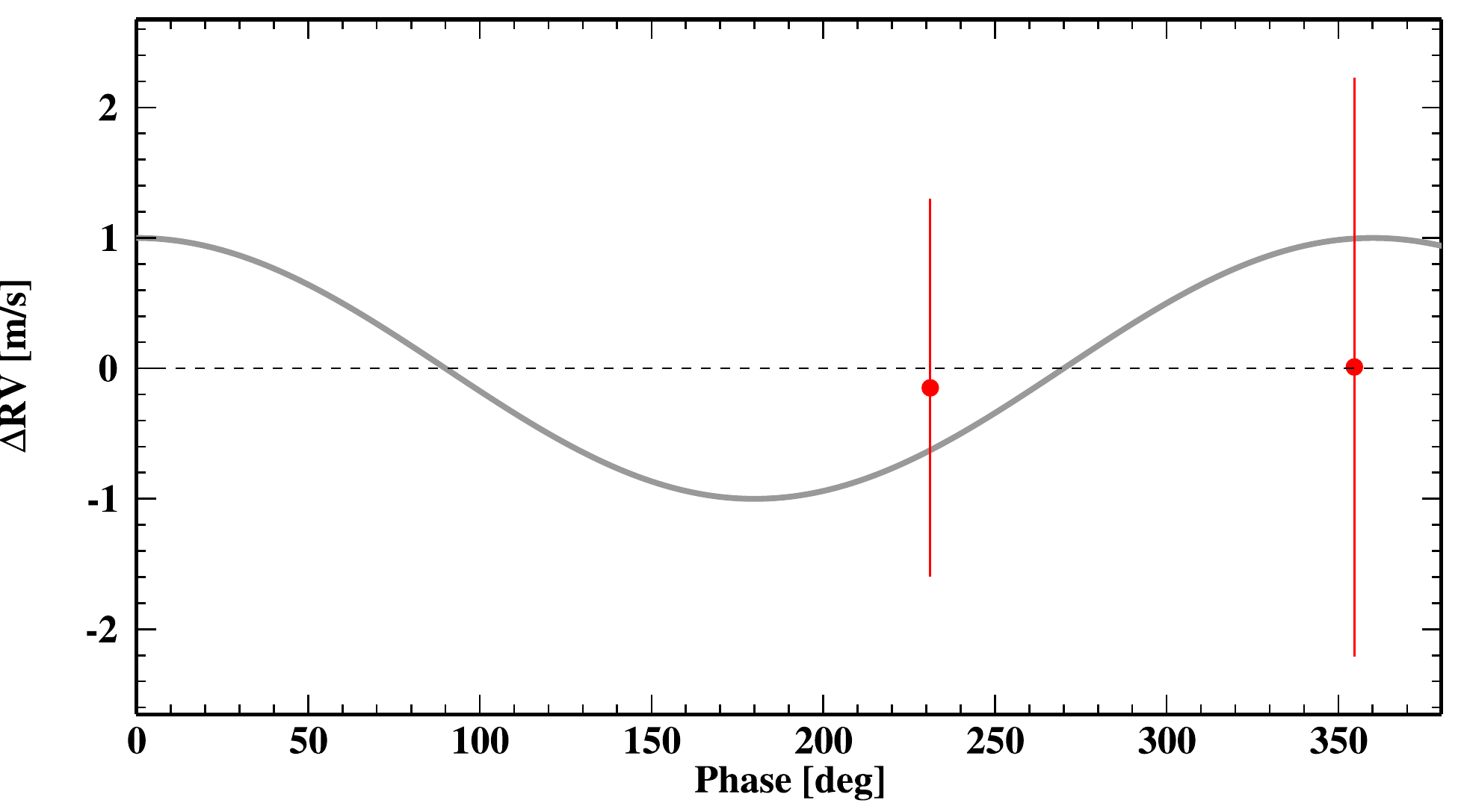}
    \caption{HARPS radial velocity measurements for NGTS candidate NG1947-4200-11647, phase-folded using the photometric period ($P=1.29297$\,d) and phase ($T_{c}=2457289.537789$).  Red circles are individual HARPS measurements and the solid line is a $K=1\,\ms$~ Keplerian orbit with $e=0$ and period and phase fixed at stated values.  The radial velocity measurements rule out this candidate as being a transiting Neptune.}
    \label{fig:harpsrvs}
\end{figure}

To confirm and determine the mass of NGTS transiting exoplanets, we use the HARPS spectrograph \citep{pepe00} on the ESO 3.6\,m telescope at La Silla Observatory,  Chile.  HARPS is a ultra-stabilised, high resolution ($R \sim 120,000$), fibre-fed echelle spectrograph designed for high precision radial velocity measurements.  HARPS is capable of sub 1\,\ms\ radial velocity precision \citep{Mayor03}, although in the case of NGTS candidates the host star magnitudes mean that we are photon limited and typically we achieve $\sim 2 - 3$\,\ms\ in a typical 45\,min exposure.  We show the example of the HARPS monitoring of NGTS candidate NG1947-4200-11647 in Fig~\ref{fig:harpsrvs}.  In this case two radial velocity epochs showed no variation at a level of $K=1\,\ms$, which when combined with the constraints from the photometric data rules out the candidate being a transiting Neptune.

For NGTS candidates around faint stars (mag > 14), and where a radial velocity precision of 30\,\ms\ is thought to be sufficient, we use the HARPS high-efficiency mode, EGGS. This gains a factor two higher throughput at the cost of increased systematics, and provides higher radial velocity precision for photon limited observations. 

\subsection{Stellar and planetary parameter estimation}
\label{sec-params}

\begin{figure}
	\includegraphics[width=8cm]{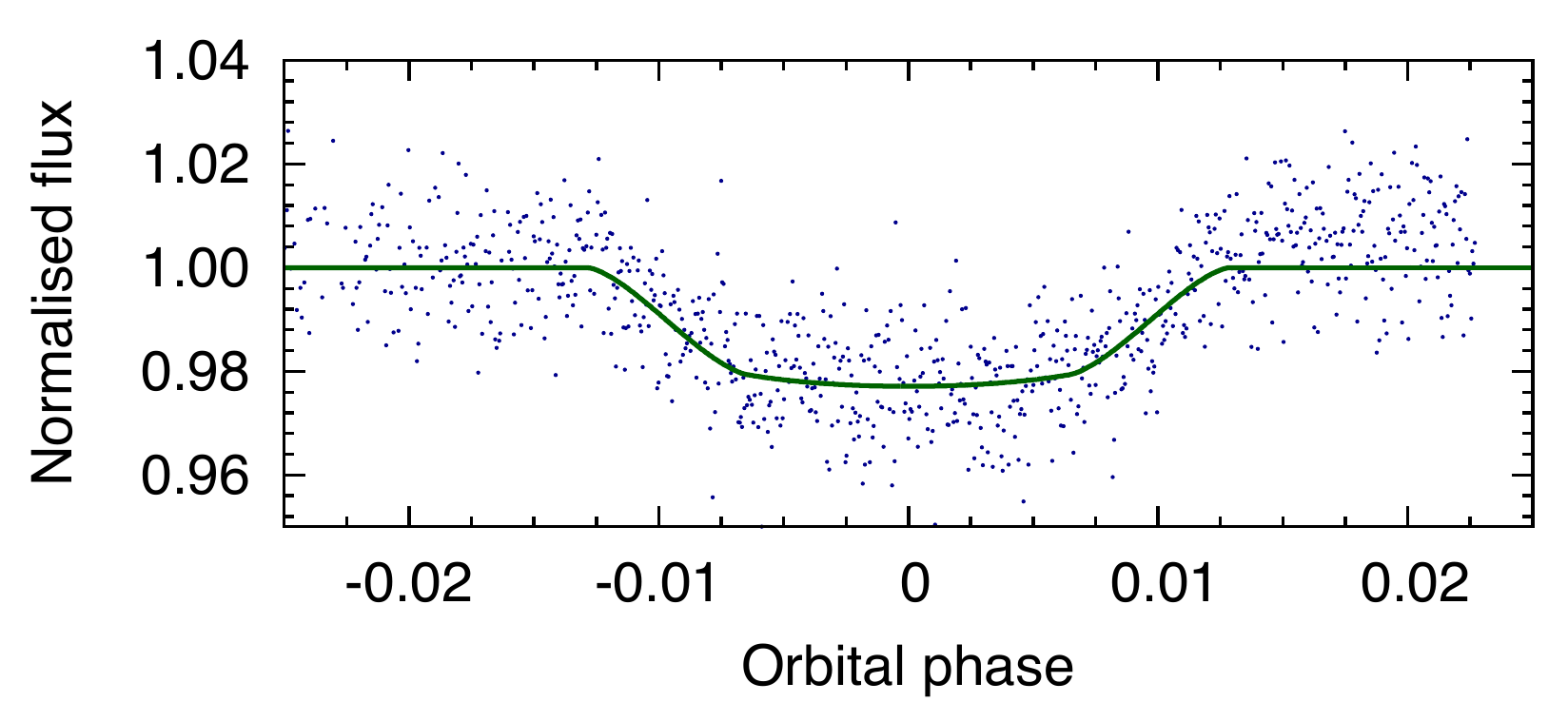}
    \includegraphics[width=8cm]{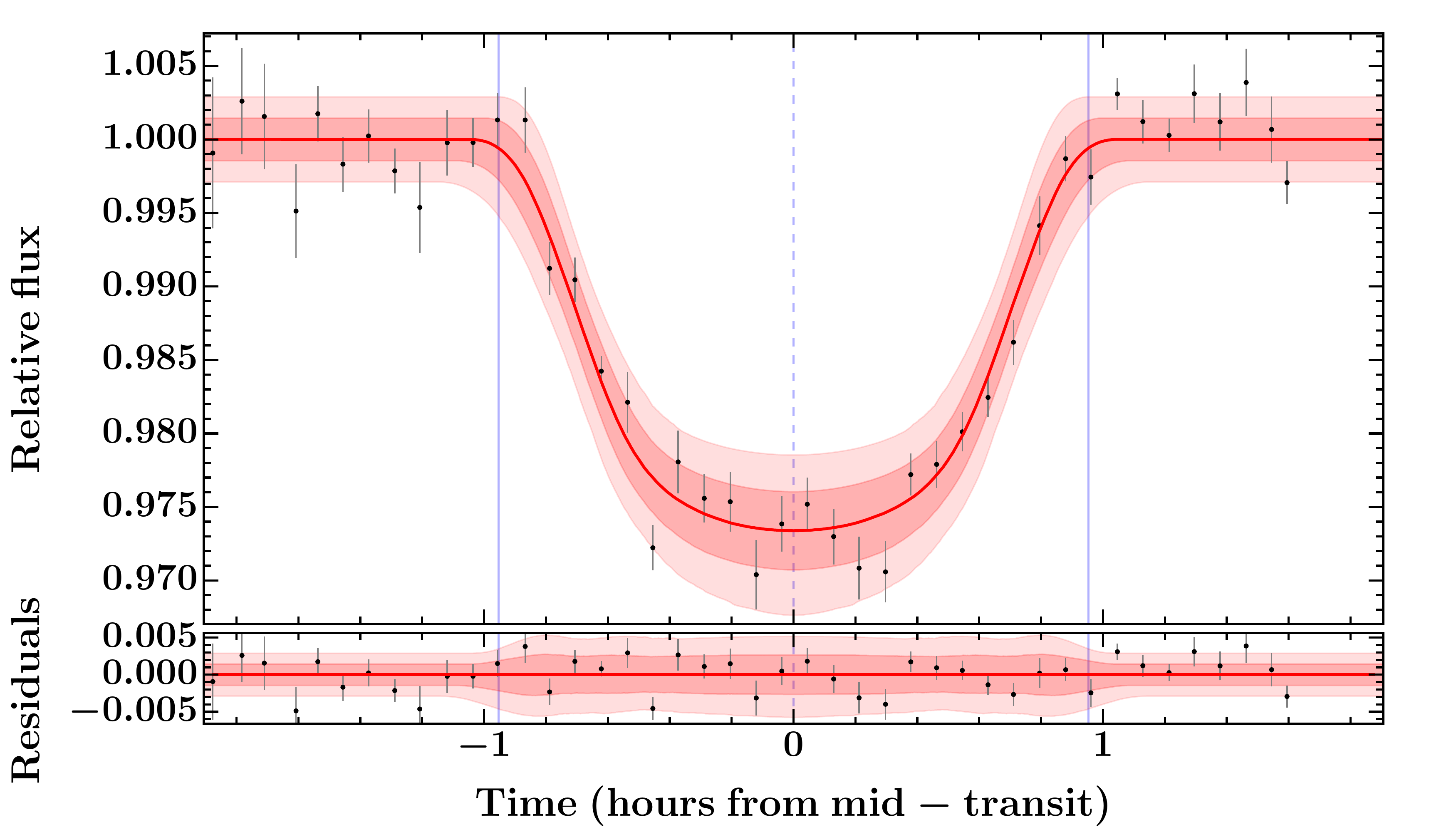}
    \caption{Model fits of an NGTS single transit observation of WASP-98b \citep{Hellier14} using TLCM (upper panel) and \gpe\ (lower panel). In the lower panel, the data was binned to 5 min cadence and \gpe\ integrated accordingly, which gives the impression of a smoother, more rounded transit.}
    \label{fig:fits}
\end{figure}

During follow up of transit candidates we fit light curves and radial velocity measurements 
with physical models 
to determine 
system parameters and estimate their uncertainties. 
We use two modelling codes, the Transit and Light Curve Modeller (TLCM) and \gpe, each of which has its own strengths. Figure~\ref{fig:fits} shows a single-transit NGTS observation of the hot Jupiter WASP-98b fitted with both TLCM and GP-EBOP. In both cases the fitted transit parameters were consistent with those from the discovery paper \citep{Hellier14}.

\label{sec:tlcm}
TLCM has been used in the discovery and modelling of exoplanets from \corot, \kepler\ and \textit{K2} \cite[e.g.][]{Csizmadia15} and it is described by Csizmadia (in preparation).  
It employs the \cite{Mandel02} model to fit the photometric transit and it uses a genetic algorithm to find the approximate global minimum followed by simulated annealing to refine the solution and determine parameter uncertainties. The photometric model includes emission from the secondary object (planet), relativistic beaming, ellipsoidal variability and reflection. TLCM can optionally include radial velocity data in the fits, including the Rossiter-McLaughlin effect, and it is applicable to eclipsing binary stars, which can be useful when vetting candidates (Sect.\,\ref{sec-vetting}). 

\gpe\ combines a Gaussian process (GP) variability model and MCMC wrapper
with a transit/eclipse model based on the EBOP family of models.
By fitting transits/eclipses simultaneously with other variability (both astrophysical and systematic) it propagates uncertainties in the variability modelling to the posterior distributions of the transit parameters.
 \gpe\ is described by \citet{Gillen17}, where it was applied to young eclipsing binary systems, and it was applied to exoplanet transits by \citet{Pepper17}. 
The transit/eclipse model accounts for reflection and ellipsoidal effects, as well as light travel time across the system \citep{Irwin11}. A quadratic limb darkening law is included following \citep{Mandel02} with limb darkening coefficients parmaeterised using the using the 
method of \citet{Kipping13} and estimated using the LDtk toolkit \citep{Parviainen15}. The GP model utilises the {\sc george} package \citep{Ambikasaran14}. When including radial velocities in the fit, \gpe\ incorporates a 
jitter term to account for stellar activity and instrument systematics. It also accounts for offsets between multiple radial velocity instruments and scales the uncertainties for each instrument individually. 

For both TLCM and \gpe, the planet properties derived from modelling transit light curves and stellar radial velocities are determined relative to the host star. We determine the host star parameters by fitting our high-resolution follow up spectra (from CORALIE, FEROS and/or HARPS) with the Spectroscopic Parameters and AtmosphEric ChemIstriEs of Stars code (SPECIES; Soto \& Jenkins, in preparation).  SPECIES allows us to determine $T_{{\rm eff}}$, $\log{g}$, metallicity, $v\sin{i}$, and micro and macroturbulence, along with the mass, radius, luminosity, and age of the star. The code is designed to run in an automated fashion, dealing with correlated parameters on the fly, and providing robust estimates of parameter uncertainties.  
Where the combined high-resolution spectra have sufficient signal-to-noise, 
SPECIES also delivers a further twelve atomic elemental abundances.
For template spectra of S/N$>$50,
our abundances can be constrained to better than 0.1\,dex, with the best estimates being made for small planets because they require more intense radial velocity follow up. 

The red-sensitive bandpass of NGTS allows us to probe the nearby early M-dwarf population (Sects.\,\ref{sec-goals}\,\&\,\ref{sec-ngts}) but determining stellar parameters for M-dwarfs is harder than for FGK stars due to uncertainties in stellar atmosphere models at low temperatures 
where molecules form and line lists are incomplete. In these cases we adopt 
empirical M-dwarf relations \citep{Mann15,Benedict16}
along with 
SED modelling 
(see Sect.\,\ref{sec-sed}).
Typically, we determine $T_{\rm eff}$, mass and radius from SED fitting using initial estimates from \citet{Mann15} as priors. We then use our SED-derived distance estimate, along with broadband magnitudes, to refine the final mass using \citet{Benedict16} (and checking for consistency with Mann et al.).

\begin{figure}
\begin{center}
\includegraphics[width=1.0\linewidth]{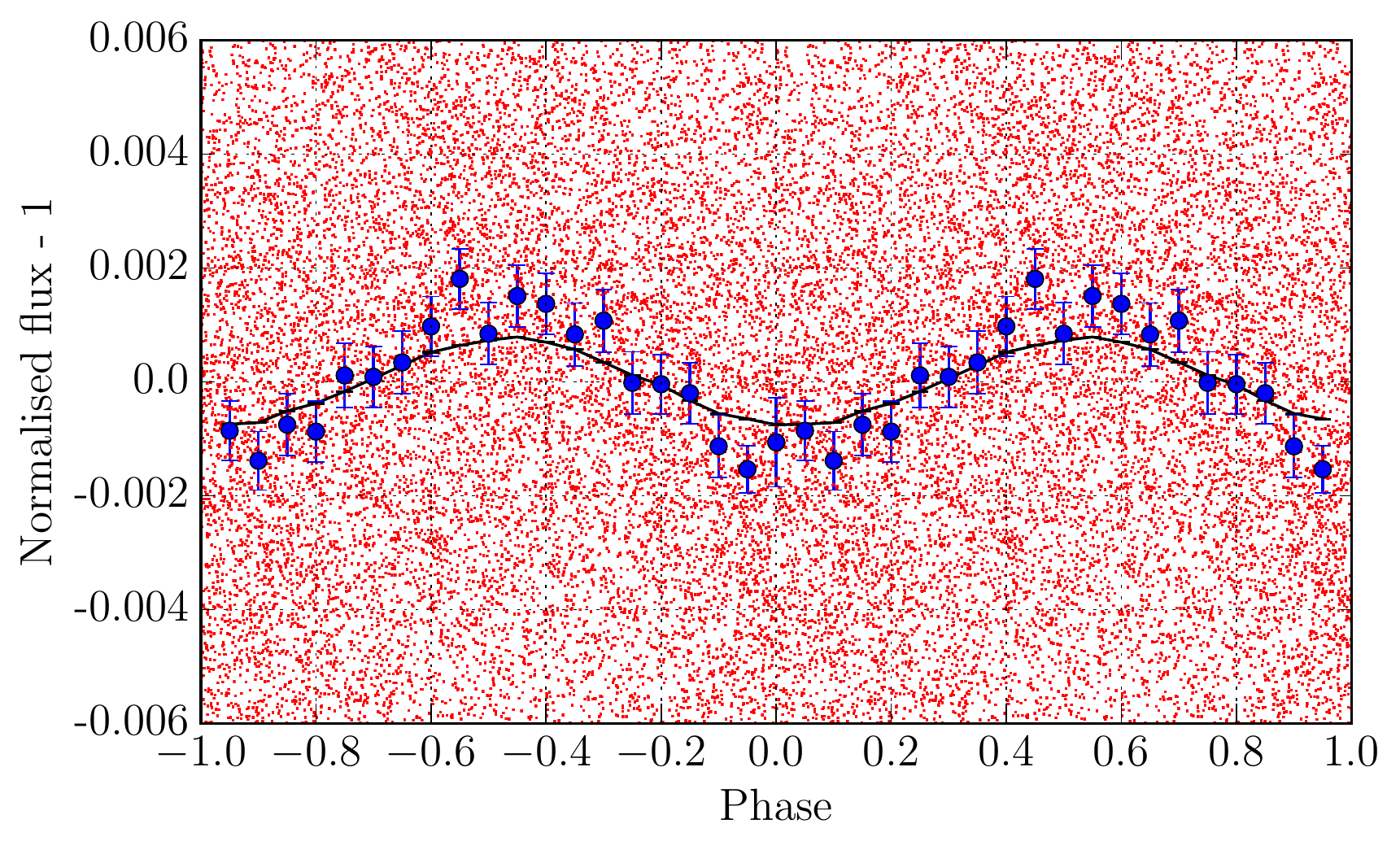}
\caption[A 1 mmag delta scuti recorded with NGTS]{A 1 mmag delta scuti detected in the primary \kepler\ field with the NGTS prototype telescope unit in Geneva.
The red points are the individual photometric measurements in 10\,s exposures, while the blue points show how the photometry bins down in phase to provide a clear detection of the 1\,mmag periodic signal. The solid black line is the folded \kepler\ light curve for this object.
}
\label{fig-delta_scuti}
\end{center}
\end{figure}

\section{Photometric performance of NGTS}

\subsection{Geneva testing}
During the commissioning phase of the NGTS project we assembled and tested one of the telescope units on the roof of the Geneva Observatory. While our primary aim was to test the integration and performance of hardware and software components, this phase also provided the opportunity to demonstrate the potential for high precision photometry with NGTS telescopes by observing well characterised variable stars in the \kepler\ field. Despite relatively poor observing conditions we made photometric measurements with 10\,s exposures across parts of seventeen separate nights between 4 June 2013 and 2 August 2013. These data provided the test-bed for the development of our data reduction and analysis pipelines described in Sects.\,\ref{sec-reduction}\,\&\,\ref{sec-analysis}. 

An example result from the Geneva testing is shown in Fig.\,\ref{fig-delta_scuti}. This is the binned (blue) and unbinned (red) phase folded NGTS measurements of KIC 11497012, which is a $\delta$ Scuti star detected in the \kepler\ survey \citep{Uytterhoeven11}.  The solid black line shows the folded \kepler\ light curve of the stellar pulsations, which have an amplitude of only 1\,mmag on a period of one hour. It can be seen that the binned NGTS light curve is a close match to \kepler, demonstrating that the individual NGTS data points bin down to high precision measurements. The signal is also independently detected 
with high significance in 
in the unbinned data using 
a Lomb-Scargle periodogram. The slightly larger amplitude detected with NGTS probably reflects our different bandpass, which is optimised for red light (Sect.\,\ref{sec-ngts} \& Fig.\,\ref{fig-qe}). 

\begin{figure}
\begin{center}
\includegraphics[width=1.0\linewidth,height=0.8\linewidth]{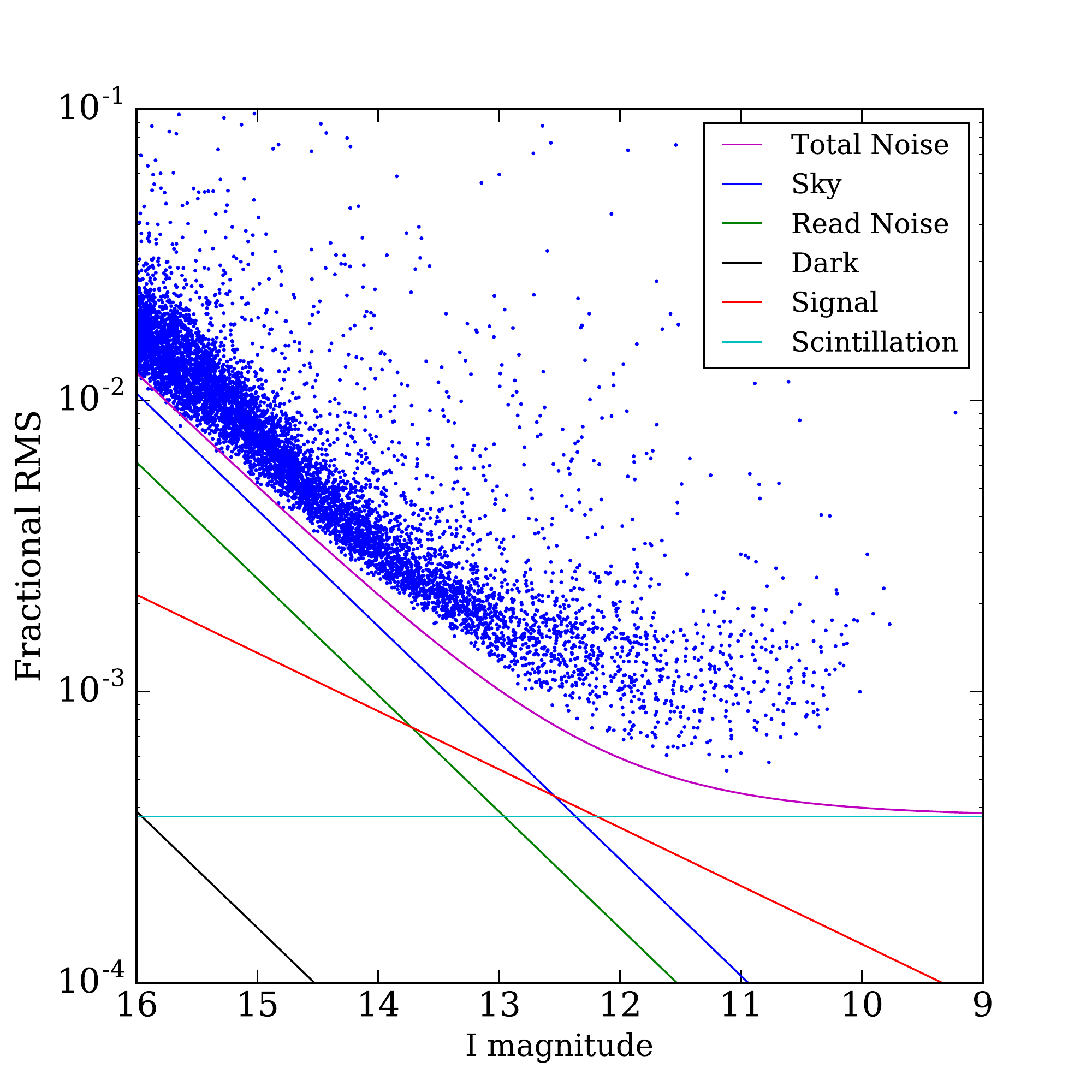}
\caption{Fractional RMS noise in detrended NGTS light curves plotted as a function of stellar brightness for one of our completed survey fields. The data span 156 nights with a total of 695 hours of high-quality photometric monitoring at 12\,s cadence (208,500 images). For this figure the data have been binned to exposure times of 1\,h. 
}
\label{fig-frms}
\end{center}
\end{figure}

\begin{figure}
\begin{center}
\includegraphics[width=1.0\linewidth]{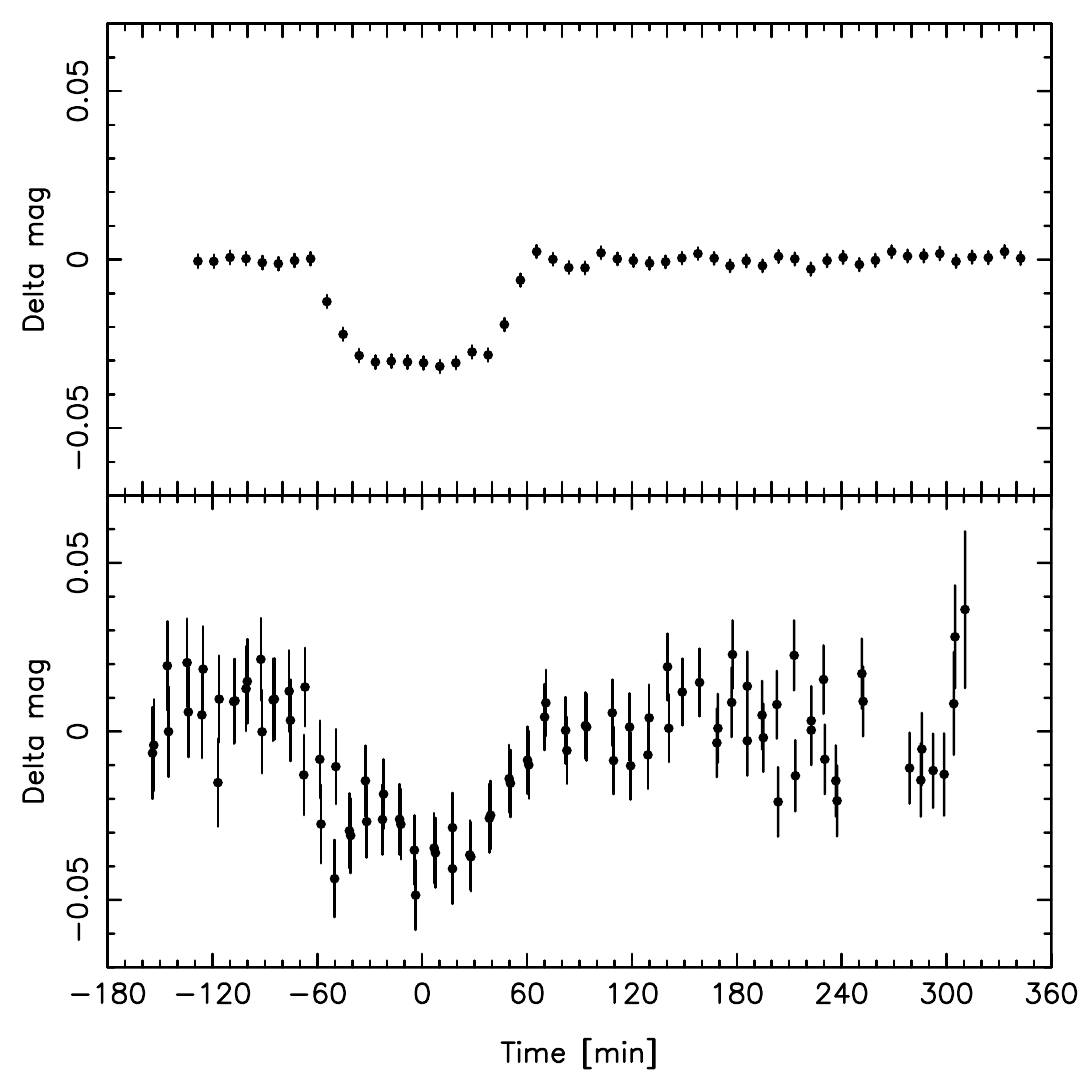}
\caption[WASP-4b with NGTS]{Single transit observations of the hot Jupiter WASP-4b with one NGTS telescope unit (top) and WASP (bottom). It can be seen that a Jupiter-sized exoplanet can be identified in a single transit with NGTS.
}
\label{fig-wasp4}
\end{center}
\end{figure}

\subsection{Full instrument at Paranal}
As summarised in Fig.\,\ref{fig-stats}, the NGTS survey began with four telescope units in September 2015. A number of full survey fields have since been completed, and in Fig.\,\ref{fig-frms} we show a summary of the noise properties of one of these completed survey fields. The data summarised here were taken at 12\,s cadence across 156 nights with a total exposure time of 579\,h (208,500 images with 10\,s exposures). We carried out photometry of 8504 stars with I band magnitudes brighter than 16, and passed the data through the reduction and detrending pipelines described in Sects.\,\ref{sec-reduction}\,\&\,\ref{sec-analysis}. While we continue to refine our pipelines, particularly with regard to precise background subtraction and flat fielding, we are encouraged by the generally close correspondence of data with our noise model. For many stars the fractional RMS noise is below 1\,mmag (for data binned to 1\,hour exposure) which we believe is the highest precision ever achieved in a wide-field ground-based sky survey. Inspection of individual light curves shows that most stars lying substantially above the noise model are genuine variables. 

For stars fainter than the scintillation limit at $I\sim12.5$, Fig.\,\ref{fig-frms} shows that the photometric precision of NGTS is comparable to that of \tess\ \citep{Ricker15}. Combining data from the two instruments, together with the flexible scheduling of NGTS, therefore has the potential to find planets that would not be detected by either instrument individually. 
This will be of particular value for M-dwarf host stars, which tend to be relatively faint, and for longer orbital periods where a long duration stare and/or flexible scheduling of transit observations is required.

In the top panel of Fig\,\ref{fig-wasp4} we plot a portion of the NGTS light curve of a known transiting exoplanet, WASP-4b \citep{Wilson08}. And we compare it with a single transit from the WASP discovery data (lower panel). With NGTS precision, it can be seen that this hot Jupiter is readily detected in a single transit. Indeed the quality of our data is comparable with that attained by specialised follow-up using much larger telescopes \citep[e.g.][]{Gillon09,Winn09,Nikolov12}. As well as demonstrating the photometric precision of our individual light curves, these data illustrate how NGTS is capable of single transit detection of long-period giant planets. 

\subsection{Transit detection efficiency}
\label{sec:det_eff}
To quantify the detection capability of NGTS, and to hone our detrending procedure (Sect.\,\ref{sec-detrend}), we developed a code to generate realistic transit signals and inject them into real NGTS light curves. We run our standard transit detection algorithms on these signals (Sect.\,\ref{sec-transits}) in order to measure the recovery rate as a function of exoplanet size and orbital period as well as stellar type and brightness. The synthetic transit signals are injected into raw light curves, and the detrending algorithms run afterwards (Sect.\,\ref{sec-detrend}), in order to account for transit signals that are weakened or removed by light curve detrending. 

\begin{figure}
	\includegraphics[width=8.4cm]{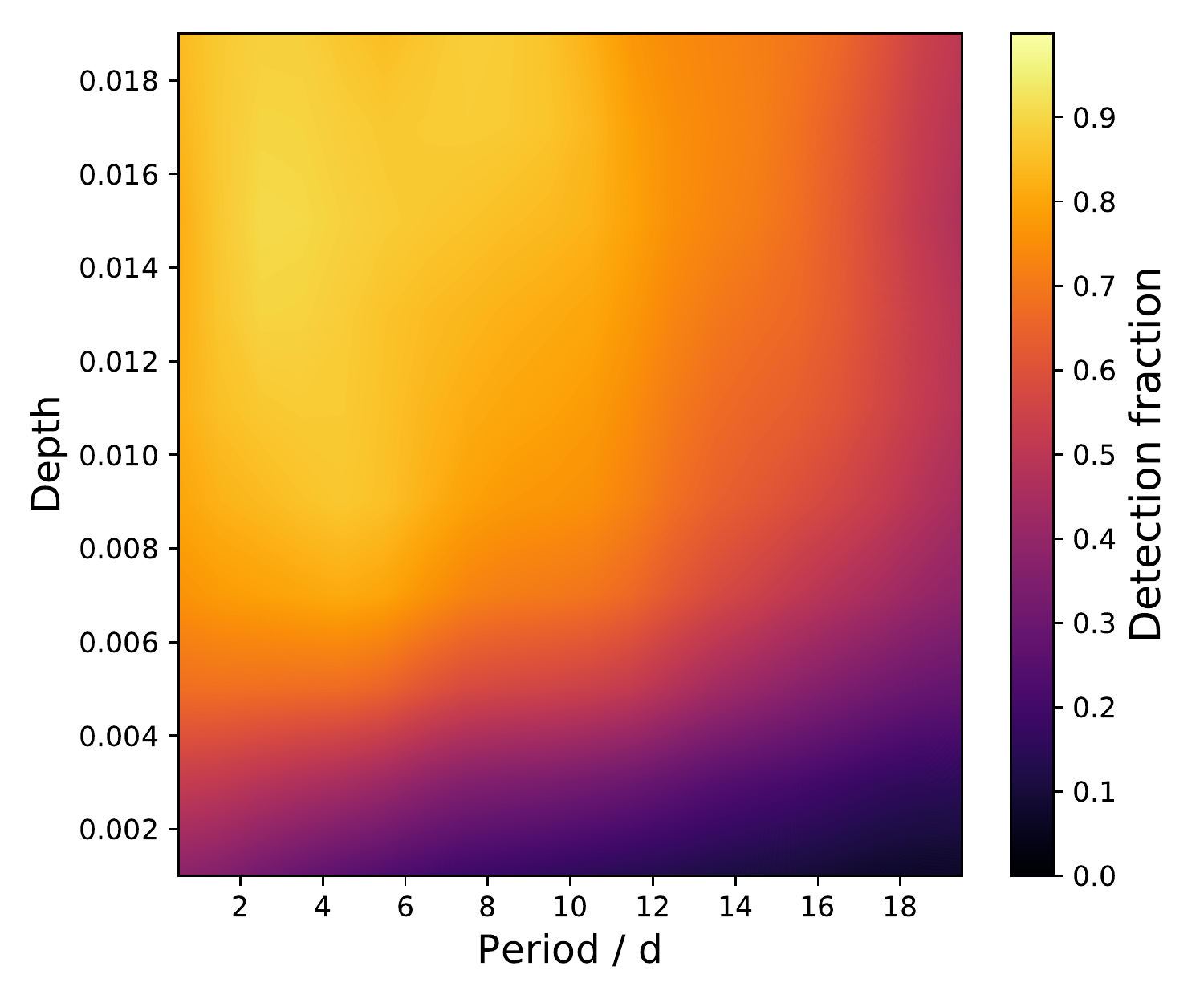}
    \caption{Transit detection efficiency as a function of transit depth and orbital period for a typical NGTS field. Around 216\,000 images were taken of this field in good photometric conditions over 149 nights, during a season spanning 225 nights. Synthetic transits were injected into the light curves of objects with NGTS magnitudes between 8 and 13, and the {\sc Orion} BLS code was used to detect them. A Gaussian filter was used to smooth the resulting density plot.}
    \label{fig-syn}
\end{figure}

\begin{figure}
\begin{center}
\includegraphics[width=1.0\linewidth]{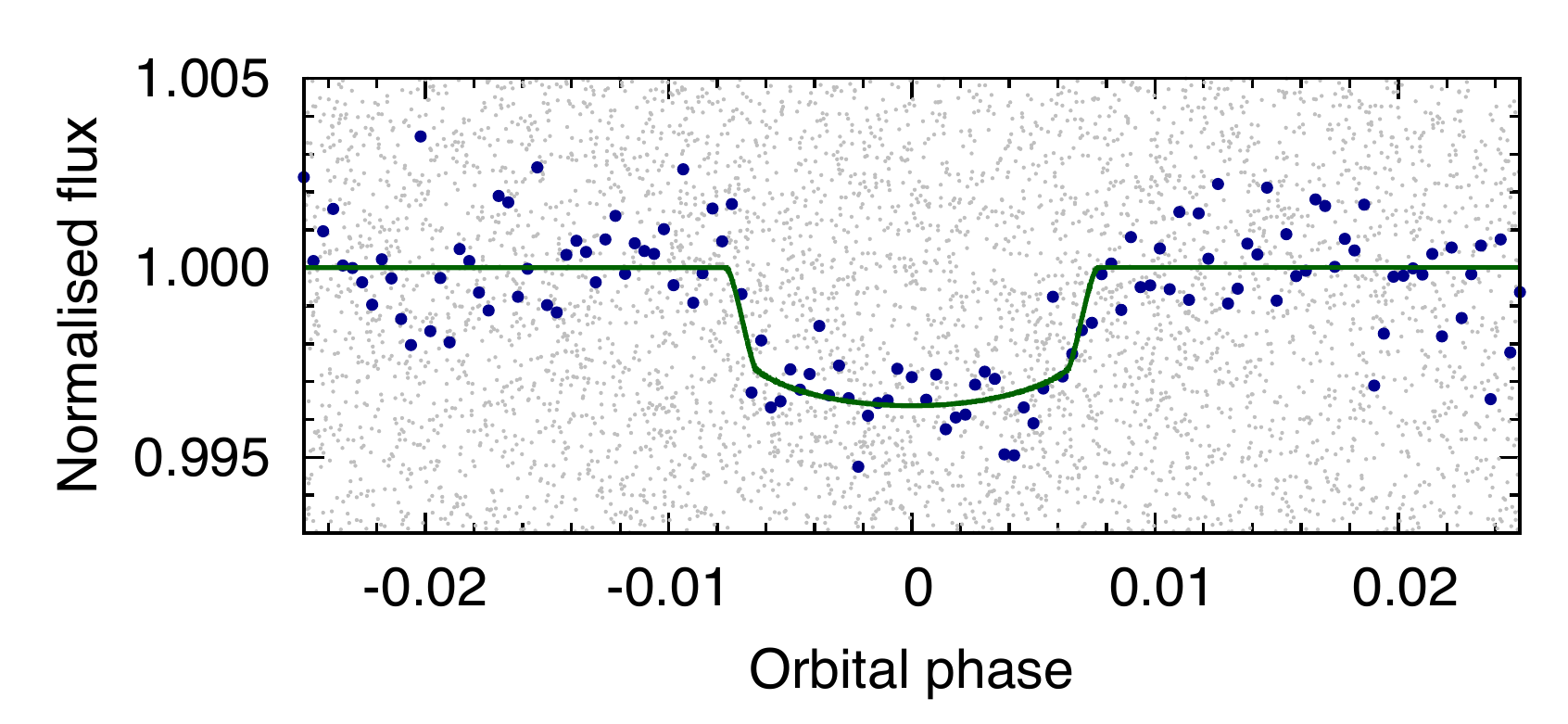}
\caption{Recovery of one of the artificial transit signals used to measure the NGTS transit detection efficiency in Sect.\,\ref{sec:det_eff} (plotted in Fig.\,\ref{fig-syn}). In this case a transit signal corresponding to a planet with radius 1.17 $\mathrm{R_{Nep}}$ and orbital period 3.53~d was injected into the light curve of a randomly selected $R = 13.30$ star, whose radius we estimate to be 0.69 $\mathrm{R_\odot}$. The grey points are the unbinned NGTS light curve, and the blue points are the same data, binned to 0.0004 in phase ($\approx 2$~min). The green curve is a TLCM fit to the data (Section~\ref{sec:tlcm}).
}
\label{fig:inj}
\end{center}
\end{figure}

To quantify the detection efficiency of our current observing strategy and data analysis methods we 
randomly selected 20\,000 light curves of stars brighter than $I=13$ from a typical, well-observed field.
Orbital periods were randomly drawn from a uniform distribution, $0.35 \leq P \leq 20.0$~d, and the squared ratio of the planetary-to-stellar radii from a distribution, 
$0.001 \leq \left( R_\mathrm{p}/R_* \right) ^2 \leq 0.02$.
Stellar radii and masses were estimated from the $J-H$ colour, using the relations of \cite{Collier-cameron07}. Random (transiting) inclination angles were chosen and orbits were assumed to be circular. The transit models were generated using the routines of \cite{Mandel02}, employing the non-linear limb-darkening law of \cite{Claret2000}. The method is described in more detail by \citet{Walker13} and Walker et al.\ (in preparation).

To compute the detection efficiency we used the {\sc Orion} transit detection code (Sect.\,\ref{sec:orion}) and applied our standard detection thresholds. These require a signal to be one of the five strongest periodic signals detected by {\sc Orion} and to have a Signal Detection Efficiency \citep[SDE;][]{Kovacs02} of at least 5. An injected signal is considered detected where the recovered period is either the injected period, $P_\mathrm{inj}$, or one of the harmonics at $P_\mathrm{inj} / 2$ or $2 P_\mathrm{inj}$ (with a tolerance of one part in a thousand). 
To produce the detection map shown in Fig.\,\ref{fig-syn}, we bin our detections using a $10 \times 20$ grid in depth -- period space, and smooth the resulting density plot for display purposes using a Gaussian filter.

Figure\,\ref{fig-syn} shows that with our current observing strategy and detrending algorithms the NGTS system is capable of detecting 
transit signals as shallow as 0.1 per cent at short orbital periods, including most signals deeper than 0.2 per cent. This reflects the low level of red noise in the NGTS photometry. Figure\,\ref{fig-syn} also shows that most signals deeper than 0.5 per cent are detected even at longer periods of up to 20\,d. This is remarkable for a ground-based experiment at a single longitude, and it reflects the good weather conditions at Paranal and our excellent single-transit precision.

To further illustrate the ability of NGTS to detect shallow transit signals associated with Neptune-sized planets, Fig.\,\ref{fig:inj} shows the recovery of one of the synthetic transit signals.
The star in question has $R = 13.30$, and $J - H = 0.777$, from which we inferred a radius of 0.69~$\mathrm{R_\odot}$. In this case a transit signal with a depth of 0.36 per cent was injected, corresponding to a planet 1.17 times the size of Neptune, with an orbital period of 
3.53\,d.

The strongest periodic signal detected by {\sc Orion} for this object has a period within 12~s of the synthetic orbital period, $P_\mathrm{inj}$. Each of the four remaining signals detected by {\sc Orion} are harmonics of this period (at close to $2 P_\mathrm{inj}$ and $4 P_\mathrm{inj}$). We modelled the light curve using TLCM (Section~\ref{sec:tlcm}), assuming a circular orbit, and find a best-fitting ratio of planetary to stellar radius that is consistent with the injected signal at the 1-$\sigma$ level. The best fitting model is overlaid in Fig.\,\ref{fig:inj}.

\section{Summary and outlook}
We have described the Next Generation Transit Survey (NGTS) and our scientific goals to detect Super-Earth and Neptune-sized exoplanets from the ground. The NGTS facility has been designed for high precision photometric measurements over a wide field of view, and it is optimised both for transiting exoplanet searches and for 
follow up of exoplanet candidates from space-based instruments such as \textit{K2}, \tess, \gaia\ and \plato.  The twelve independently-pointable telescopes allow for efficient monitoring of targets spread across the observable sky. 

NGTS has been operating routinely since April 2016 at the ESO Paranal Observatory, Chile.
The high photometric precision --- better than 1\,mmag for most stars brighter than 12th magnitude --- 
is made possible by 
the quality of the Paranal site,
precise autoguiding, 
and an instrument design that ensures stable pointing and image shape. This stability also allows the application of a centroiding method that efficiently identifies false positive transits. Transit injections show that most signals deeper than 0.2 per cent are detected at short orbital periods and most signals deeper than 0.5 per cent are detected even at periods as long as 20\,d.

The photometric precision of NGTS is comparable to that of \tess\ for stars fainter than the scintillation limit ($I\sim12.5$), and flexibly-scheduled NGTS photometry of single-transit candidates from \tess\ will extend planet discoveries to longer orbital periods than can be detected in the standard 27\,d \tess\ dwell time.  
NGTS photometry will also test \tess\ candidates for false positives due to blended eclipsing binaries and provide precise ephemerides for atmospheric characterisation with flagship facilities such as \jwst\ and E-ELT.

The NGTS exoplanet survey is under way and the first NGTS exoplanets have already been confirmed by radial velocity observations with HARPS 
\citep[e.g.][]{Bayliss17}.
NGTS data are also being used for a wide range of variable star studies. 
When \tess\ data become available, a joint analysis with archival NGTS survey data will allow searches for shallow transit signals at long orbital periods that are not detectable in either instrument individually.
NGTS also stands ready to support \plato\ by characterising stellar variability and activity in advance of target selection, 
and it will be able to 
search for transits of wide-separation exoplanets with edge-on orbits detected in \gaia\ astrometry. 

NGTS data will be made publicly available through the ESO data archive. Annual data releases are planned from mid 2018, and will include raw and detrended light curves for stars brighter than $I\sim16$. We expect to a support a large user community carrying out a wide range of science projects, and we encourage potential collaborators to contact us in order to optimise the use of the NGTS for maximum scientific return.

\section*{Acknowledgements}
The capital costs of the NGTS facility were funded by 
the University of Warwick,
the University of Leicester,
Queen's University Belfast,
the University of Geneva,
the Deutsches Zentrum f\" ur Luft- und Raumfahrt e.V. (DLR; under the `Gro\ss investition GI-NGTS'),
the University of Cambridge
and the UK Science and Technology Facilities Council (STFC; project reference ST/M001962/1). The facility is operated by the consortium institutes with support from STFC (also project ST/M001962/1). 
We are grateful to ESO for providing access to the Paranal site as well as generous in-kind support. The research leading to these results has received funding from the European Research Council (ERC) under the European Union's Seventh Framework Programme (FP/2007-2013) / ERC Grant Agreement n. 320964 (WDTracer). 
The contributions at the University of Warwick by PJW, RGW, DLP, FF, DA, BTG and TL have been supported by STFC through consolidated grants ST/L000733/1 and ST/P000495/1. 
TL was also supported by STFC studentship 1226157. 
EF is funded by the Qatar National Research Foundation (programme QNRF-NPRP-X-019-1).
MNG is supported by STFC studentship 1490409 as well as the Isaac Newton Studentship.
JSJ acknowledges support by Fondecyt grant 1161218 and partial support by CATA-Basal (PB06, CONICYT). 
AJ acknowledges support from FONDECYT project 1171208, BASAL CATA PFB-06, and by the Ministry for the Economy, Development, and Tourism's Programa Iniciativa Cient\'{i}fica Milenio through grant IC\,120009, awarded to the Millennium Institute of Astrophysics (MAS).
This research made use of Astropy, a community-developed core Python package for Astronomy \citep{Astropy13}, and NASA's Astrophysics Data System Bibliographic Services.




\bibliographystyle{mnras}
\bibliography{paper} 

\begin{thebibliography}{}
\makeatletter
\relax
\def\mn@urlcharsother{\let\do\@makeother \do\$\do\&\do\#\do\^\do\_\do\%\do\~}
\def\mn@doi{\begingroup\mn@urlcharsother \@ifnextchar [ {\mn@doi@}
  {\mn@doi@[]}}
\def\mn@doi@[#1]#2{\def\@tempa{#1}\ifx\@tempa\@empty \href
  {http://dx.doi.org/#2} {doi:#2}\else \href {http://dx.doi.org/#2} {#1}\fi
  \endgroup}
\def\mn@eprint#1#2{\mn@eprint@#1:#2::\@nil}
\def\mn@eprint@arXiv#1{\href {http://arxiv.org/abs/#1} {{\tt arXiv:#1}}}
\def\mn@eprint@dblp#1{\href {http://dblp.uni-trier.de/rec/bibtex/#1.xml}
  {dblp:#1}}
\def\mn@eprint@#1:#2:#3:#4\@nil{\def\@tempa {#1}\def\@tempb {#2}\def\@tempc
  {#3}\ifx \@tempc \@empty \let \@tempc \@tempb \let \@tempb \@tempa \fi \ifx
  \@tempb \@empty \def\@tempb {arXiv}\fi \@ifundefined
  {mn@eprint@\@tempb}{\@tempb:\@tempc}{\expandafter \expandafter \csname
  mn@eprint@\@tempb\endcsname \expandafter{\@tempc}}}

\bibitem[\protect\citeauthoryear{{Alard}}{{Alard}}{2000}]{Alard2000}
{Alard} C.,  2000, \mn@doi [\aaps] {10.1051/aas:2000214}, \href
  {http://adsabs.harvard.edu/abs/2000A%26AS..144..363A} {144, 363}

\bibitem[\protect\citeauthoryear{{Albrecht} et~al.,}{{Albrecht}
  et~al.}{2012}]{Albrecht12}
{Albrecht} S.,  et~al., 2012, \mn@doi [\apj] {10.1088/0004-637X/757/1/18},
  \href {http://adsabs.harvard.edu/abs/2012ApJ...757...18A} {757, 18}

\bibitem[\protect\citeauthoryear{{Ambikasaran}, {Foreman-Mackey}, {Greengard},
  {Hogg}  \& {O'Neil}}{{Ambikasaran} et~al.}{2014}]{Ambikasaran14}
{Ambikasaran} S.,  {Foreman-Mackey} D.,  {Greengard} L.,  {Hogg} D.~W.,
  {O'Neil} M.,  2014, preprint, \href
  {http://adsabs.harvard.edu/abs/2014arXiv1403.6015A} {} (\mn@eprint {arXiv}
  {1403.6015})

\bibitem[\protect\citeauthoryear{{Armstrong}, {Pollacco}  \&
  {Santerne}}{{Armstrong} et~al.}{2017}]{Armstrong17a}
{Armstrong} D.~J.,  {Pollacco} D.,   {Santerne} A.,  2017, \mn@doi [\mnras]
  {10.1093/mnras/stw2881}, \href
  {http://adsabs.harvard.edu/abs/2017MNRAS.465.2634A} {465, 2634}

\bibitem[\protect\citeauthoryear{{Astropy Collaboration} et~al.,}{{Astropy
  Collaboration} et~al.}{2013}]{Astropy13}
{Astropy Collaboration} et~al., 2013, \mn@doi [\aap]
  {10.1051/0004-6361/201322068}, \href
  {http://adsabs.harvard.edu/abs/2013A%26A...558A..33A} {558, A33}

\bibitem[\protect\citeauthoryear{{Auvergne} et~al.,}{{Auvergne}
  et~al.}{2009}]{Auvergne09}
{Auvergne} M.,  et~al., 2009, \mn@doi [\aap] {10.1051/0004-6361/200810860},
  \href {http://adsabs.harvard.edu/abs/2009A%26A...506..411A} {506, 411}

\bibitem[\protect\citeauthoryear{{Bakos}, {Noyes}, {Kov{\'a}cs}, {Stanek},
  {Sasselov}  \& {Domsa}}{{Bakos} et~al.}{2004}]{Bakos04}
{Bakos} G.,  {Noyes} R.~W.,  {Kov{\'a}cs} G.,  {Stanek} K.~Z.,  {Sasselov}
  D.~D.,   {Domsa} I.,  2004, \mn@doi [\pasp] {10.1086/382735}, \href
  {http://adsabs.harvard.edu/abs/2004PASP..116..266B} {116, 266}

\bibitem[\protect\citeauthoryear{{Bakos} et~al.,}{{Bakos}
  et~al.}{2010}]{Bakos10}
{Bakos} G.~{\'A}.,  et~al., 2010, \mn@doi [\apj]
  {10.1088/0004-637X/710/2/1724}, \href
  {http://adsabs.harvard.edu/abs/2010ApJ...710.1724B} {710, 1724}

\bibitem[\protect\citeauthoryear{{Bakos} et~al.,}{{Bakos} et~al.}{2013}]{HATS}
{Bakos} G.~{\'A}.,  et~al., 2013, \mn@doi [\pasp] {10.1086/669529}, \href
  {http://adsabs.harvard.edu/abs/2013PASP..125..154B} {125, 154}

\bibitem[\protect\citeauthoryear{{Baraffe}, {Chabrier}  \& {Barman}}{{Baraffe}
  et~al.}{2008}]{Baraffe08}
{Baraffe} I.,  {Chabrier} G.,   {Barman} T.,  2008, \mn@doi [\aap]
  {10.1051/0004-6361:20079321}, \href
  {http://adsabs.harvard.edu/abs/2008A%26A...482..315B} {482, 315}

\bibitem[\protect\citeauthoryear{{Batalha} et~al.,}{{Batalha}
  et~al.}{2011}]{Batalha11}
{Batalha} N.~M.,  et~al., 2011, \mn@doi [\apj] {10.1088/0004-637X/729/1/27},
  \href {http://adsabs.harvard.edu/abs/2011ApJ...729...27B} {729, 27}

\bibitem[\protect\citeauthoryear{{Bayliss} et~al.}{{Bayliss}
  et~al.}{2017}]{Bayliss17}
{Bayliss} D.,  et~al., 2017, \mnras, in press

\bibitem[\protect\citeauthoryear{{Benedict} et~al.,}{{Benedict}
  et~al.}{2016}]{Benedict16}
{Benedict} G.~F.,  et~al., 2016, \mn@doi [\aj] {10.3847/0004-6256/152/5/141},
  \href {http://adsabs.harvard.edu/abs/2016AJ....152..141B} {152, 141}

\bibitem[\protect\citeauthoryear{{Berta-Thompson} et~al.,}{{Berta-Thompson}
  et~al.}{2015}]{Berta-Thompson15}
{Berta-Thompson} Z.~K.,  et~al., 2015, \mn@doi [\nat] {10.1038/nature15762},
  \href {http://adsabs.harvard.edu/abs/2015Natur.527..204B} {527, 204}

\bibitem[\protect\citeauthoryear{{Bonfils} et~al.,}{{Bonfils}
  et~al.}{2012}]{Bonfils12}
{Bonfils} X.,  et~al., 2012, \mn@doi [\aap] {10.1051/0004-6361/201219623},
  \href {http://adsabs.harvard.edu/abs/2012A%26A...546A..27B} {546, A27}

\bibitem[\protect\citeauthoryear{{Borucki} et~al.,}{{Borucki}
  et~al.}{2010}]{Borucki10}
{Borucki} W.~J.,  et~al., 2010, \mn@doi [Science] {10.1126/science.1185402},
  \href {http://adsabs.harvard.edu/abs/2010Sci...327..977B} {327, 977}

\bibitem[\protect\citeauthoryear{{Brahm}, {Jord{\'a}n}  \& {Espinoza}}{{Brahm}
  et~al.}{2017}]{Brahm17}
{Brahm} R.,  {Jord{\'a}n} A.,   {Espinoza} N.,  2017, \mn@doi [\pasp]
  {10.1088/1538-3873/aa5455}, \href
  {http://adsabs.harvard.edu/abs/2017PASP..129c4002B} {129, 034002}

\bibitem[\protect\citeauthoryear{{Burton}, {Watson}, {Littlefair}, {Dhillon},
  {Gibson}, {Marsh}  \& {Pollacco}}{{Burton} et~al.}{2012}]{Burton12}
{Burton} J.~R.,  {Watson} C.~A.,  {Littlefair} S.~P.,  {Dhillon} V.~S.,
  {Gibson} N.~P.,  {Marsh} T.~R.,   {Pollacco} D.,  2012, \mn@doi [\apjs]
  {10.1088/0067-0049/201/2/36}, \href
  {http://adsabs.harvard.edu/abs/2012ApJS..201...36B} {201, 36}

\bibitem[\protect\citeauthoryear{{Cabrera}, {Csizmadia}, {Erikson}, {Rauer}  \&
  {Kirste}}{{Cabrera} et~al.}{2012}]{Cabrera12}
{Cabrera} J.,  {Csizmadia} S.,  {Erikson} A.,  {Rauer} H.,   {Kirste} S.,
  2012, \mn@doi [\aap] {10.1051/0004-6361/201219337}, \href
  {http://adsabs.harvard.edu/abs/2012A%26A...548A..44C} {548, A44}

\bibitem[\protect\citeauthoryear{{Calabretta} \& {Greisen}}{{Calabretta} \&
  {Greisen}}{2002}]{Calabretta02}
{Calabretta} M.~R.,  {Greisen} E.~W.,  2002, \mn@doi [\aap]
  {10.1051/0004-6361:20021327}, \href
  {http://adsabs.harvard.edu/abs/2002A%26A...395.1077C} {395, 1077}

\bibitem[\protect\citeauthoryear{{Carter} \& {Agol}}{{Carter} \&
  {Agol}}{2013}]{Carter13}
{Carter} J.~A.,  {Agol} E.,  2013, \mn@doi [\apj]
  {10.1088/0004-637X/765/2/132}, \href
  {http://adsabs.harvard.edu/abs/2013ApJ...765..132C} {765, 132}

\bibitem[\protect\citeauthoryear{{Charbonneau}, {Brown}, {Latham}  \&
  {Mayor}}{{Charbonneau} et~al.}{2000}]{Charbonneau00}
{Charbonneau} D.,  {Brown} T.~M.,  {Latham} D.~W.,   {Mayor} M.,  2000, \mn@doi
  [\apjl] {10.1086/312457}, \href
  {http://adsabs.harvard.edu/abs/2000ApJ...529L..45C} {529, L45}

\bibitem[\protect\citeauthoryear{{Charbonneau}, {Brown}, {Noyes}  \&
  {Gilliland}}{{Charbonneau} et~al.}{2002}]{Charbonneau02}
{Charbonneau} D.,  {Brown} T.~M.,  {Noyes} R.~W.,   {Gilliland} R.~L.,  2002,
  \mn@doi [\apj] {10.1086/338770}, \href
  {http://adsabs.harvard.edu/abs/2002ApJ...568..377C} {568, 377}

\bibitem[\protect\citeauthoryear{{Charbonneau} et~al.,}{{Charbonneau}
  et~al.}{2005}]{Charbonneau05}
{Charbonneau} D.,  et~al., 2005, \mn@doi [\apj] {10.1086/429991}, \href
  {http://adsabs.harvard.edu/abs/2005ApJ...626..523C} {626, 523}

\bibitem[\protect\citeauthoryear{{Charbonneau} et~al.,}{{Charbonneau}
  et~al.}{2009}]{Charbonneau09}
{Charbonneau} D.,  et~al., 2009, \mn@doi [\nat] {10.1038/nature08679}, \href
  {http://adsabs.harvard.edu/abs/2009Natur.462..891C} {462, 891}

\bibitem[\protect\citeauthoryear{{Chazelas} et~al.,}{{Chazelas}
  et~al.}{2012}]{Chazelas12}
{Chazelas} B.,  et~al., 2012, in Ground-based and Airborne Telescopes IV. p.
  84440E, \mn@doi{10.1117/12.925755}

\bibitem[\protect\citeauthoryear{{Chromey} \& {Hasselbacher}}{{Chromey} \&
  {Hasselbacher}}{1996}]{Chromey96}
{Chromey} F.~R.,  {Hasselbacher} D.~A.,  1996, \mn@doi [\pasp]
  {10.1086/133817}, \href {http://adsabs.harvard.edu/abs/1996PASP..108..944C}
  {108, 944}

\bibitem[\protect\citeauthoryear{{Claret}}{{Claret}}{2000}]{Claret2000}
{Claret} A.,  2000, \aap, \href
  {http://adsabs.harvard.edu/abs/2000A%26A...363.1081C} {363, 1081}

\bibitem[\protect\citeauthoryear{{Collier Cameron} et~al.,}{{Collier Cameron}
  et~al.}{2006}]{Cameron06}
{Collier Cameron} A.,  et~al., 2006, \mn@doi [\mnras]
  {10.1111/j.1365-2966.2006.11074.x}, \href
  {http://adsabs.harvard.edu/abs/2006MNRAS.373..799C} {373, 799}

\bibitem[\protect\citeauthoryear{{Collier Cameron} et~al.,}{{Collier Cameron}
  et~al.}{2007}]{Collier-cameron07}
{Collier Cameron} A.,  et~al., 2007, \mn@doi [\mnras]
  {10.1111/j.1365-2966.2007.12195.x}, \href
  {http://adsabs.harvard.edu/abs/2007MNRAS.380.1230C} {380, 1230}

\bibitem[\protect\citeauthoryear{{Csizmadia} et~al.,}{{Csizmadia}
  et~al.}{2015}]{Csizmadia15}
{Csizmadia} S.,  et~al., 2015, \mn@doi [\aap] {10.1051/0004-6361/201526763},
  \href {http://adsabs.harvard.edu/abs/2015A%26A...584A..13C} {584, A13}

\bibitem[\protect\citeauthoryear{{Cutri} \& {et al.}}{{Cutri} \& {et
  al.}}{2014}]{Cutri14}
{Cutri} R.~M.,  {et al.} 2014, VizieR Online Data Catalog, \href
  {http://cdsads.u-strasbg.fr/abs/2014yCat.2328....0C} {2328}

\bibitem[\protect\citeauthoryear{{Deming}, {Seager}, {Richardson}  \&
  {Harrington}}{{Deming} et~al.}{2005}]{Deming05}
{Deming} D.,  {Seager} S.,  {Richardson} L.~J.,   {Harrington} J.,  2005,
  \mn@doi [\nat] {10.1038/nature03507}, \href
  {http://adsabs.harvard.edu/abs/2005Natur.434..740D} {434, 740}

\bibitem[\protect\citeauthoryear{{Dittmann} et~al.,}{{Dittmann}
  et~al.}{2017}]{Dittmann17}
{Dittmann} J.~A.,  et~al., 2017, \mn@doi [\nat] {10.1038/nature22055}, \href
  {http://adsabs.harvard.edu/abs/2017Natur.544..333D} {544, 333}

\bibitem[\protect\citeauthoryear{{Doyle} et~al.,}{{Doyle}
  et~al.}{2011}]{Doyle11}
{Doyle} L.~R.,  et~al., 2011, \mn@doi [Science] {10.1126/science.1210923},
  \href {http://adsabs.harvard.edu/abs/2011Sci...333.1602D} {333, 1602}

\bibitem[\protect\citeauthoryear{{Dragomir} et~al.,}{{Dragomir}
  et~al.}{2013}]{Dragomir13}
{Dragomir} D.,  et~al., 2013, \mn@doi [\apjl] {10.1088/2041-8205/772/1/L2},
  \href {http://adsabs.harvard.edu/abs/2013ApJ...772L...2D} {772, L2}

\bibitem[\protect\citeauthoryear{{Dravins}, {Lindegren}, {Mezey}  \&
  {Young}}{{Dravins} et~al.}{1998}]{Dravins98}
{Dravins} D.,  {Lindegren} L.,  {Mezey} E.,   {Young} A.~T.,  1998, \mn@doi
  [\pasp] {10.1086/316161}, \href
  {http://adsabs.harvard.edu/abs/1998PASP..110..610D} {110, 610}

\bibitem[\protect\citeauthoryear{{Ehrenreich} et~al.,}{{Ehrenreich}
  et~al.}{2015}]{Ehrenreich15}
{Ehrenreich} D.,  et~al., 2015, \mn@doi [\nat] {10.1038/nature14501}, \href
  {http://adsabs.harvard.edu/abs/2015Natur.522..459E} {522, 459}

\bibitem[\protect\citeauthoryear{{Fitzpatrick}}{{Fitzpatrick}}{1999}]{Fitzpatrick99}
{Fitzpatrick} E.~L.,  1999, \mn@doi [\pasp] {10.1086/316293}, \href
  {http://adsabs.harvard.edu/abs/1999PASP..111...63F} {111, 63}

\bibitem[\protect\citeauthoryear{{Gaia Collaboration} et~al.,}{{Gaia
  Collaboration} et~al.}{2016}]{Gaia2016}
{Gaia Collaboration} et~al., 2016, \mn@doi [\aap]
  {10.1051/0004-6361/201629512}, \href
  {http://adsabs.harvard.edu/abs/2016A%26A...595A...2G} {595, A2}

\bibitem[\protect\citeauthoryear{{Gillen}, {Hillenbrand}, {David}, {Aigrain},
  {Rebull}, {Stauffer}, {Cody}  \& {Queloz}}{{Gillen} et~al.}{2017}]{Gillen17}
{Gillen} E.,  {Hillenbrand} L.~A.,  {David} T.~J.,  {Aigrain} S.,  {Rebull} L.,
   {Stauffer} J.,  {Cody} A.~M.,   {Queloz} D.,  2017, \mn@doi [\apj]
  {10.3847/1538-4357/849/1/11}, \href
  {http://adsabs.harvard.edu/abs/2017arXiv170603084G} {849, 11}

\bibitem[\protect\citeauthoryear{{Gillon} et~al.,}{{Gillon}
  et~al.}{2007}]{Gillon07}
{Gillon} M.,  et~al., 2007, \mn@doi [\aap] {10.1051/0004-6361:20077799}, \href
  {http://adsabs.harvard.edu/abs/2007A%26A...472L..13G} {472, L13}

\bibitem[\protect\citeauthoryear{{Gillon} et~al.,}{{Gillon}
  et~al.}{2009}]{Gillon09}
{Gillon} M.,  et~al., 2009, \mn@doi [\aap] {10.1051/0004-6361:200810929}, \href
  {http://adsabs.harvard.edu/abs/2009A%26A...496..259G} {496, 259}

\bibitem[\protect\citeauthoryear{{Gillon} et~al.,}{{Gillon}
  et~al.}{2016}]{Gillon16}
{Gillon} M.,  et~al., 2016, \mn@doi [\nat] {10.1038/nature17448}, \href
  {http://adsabs.harvard.edu/abs/2016Natur.533..221G} {533, 221}

\bibitem[\protect\citeauthoryear{{Gillon} et~al.,}{{Gillon}
  et~al.}{2017}]{Gillon17}
{Gillon} M.,  et~al., 2017, \mn@doi [\nat] {10.1038/nature21360}, \href
  {http://adsabs.harvard.edu/abs/2017Natur.542..456G} {542, 456}

\bibitem[\protect\citeauthoryear{{Gray} \& {Corbally}}{{Gray} \&
  {Corbally}}{2009}]{Gray09}
{Gray} R.~O.,  {Corbally} J. C.,  2009, {Stellar Spectral Classification}.
{Princeton University Press}

\bibitem[\protect\citeauthoryear{{Greisen} \& {Calabretta}}{{Greisen} \&
  {Calabretta}}{2002}]{Greisen02}
{Greisen} E.~W.,  {Calabretta} M.~R.,  2002, \mn@doi [\aap]
  {10.1051/0004-6361:20021326}, \href
  {http://adsabs.harvard.edu/abs/2002A%26A...395.1061G} {395, 1061}

\bibitem[\protect\citeauthoryear{{G{\"u}nther}, {Queloz}, {Demory}  \&
  {Bouchy}}{{G{\"u}nther} et~al.}{2017a}]{Guenther17}
{G{\"u}nther} M.~N.,  {Queloz} D.,  {Demory} B.-O.,   {Bouchy} F.,  2017a,
  \mn@doi [\mnras] {10.1093/mnras/stw2908}, \href
  {http://adsabs.harvard.edu/abs/2017MNRAS.465.3379G} {465, 3379}

\bibitem[\protect\citeauthoryear{{G{\"u}nther} et~al.,}{{G{\"u}nther}
  et~al.}{2017b}]{Guenther17a}
{G{\"u}nther} M.~N.,  et~al., 2017b, \mn@doi [\mnras] {10.1093/mnras/stx1920},
  472, 295

\bibitem[\protect\citeauthoryear{{Hellier} et~al.,}{{Hellier}
  et~al.}{2014}]{Hellier14}
{Hellier} C.,  et~al., 2014, \mn@doi [\mnras] {10.1093/mnras/stu410}, \href
  {http://adsabs.harvard.edu/abs/2014MNRAS.440.1982H} {440, 1982}

\bibitem[\protect\citeauthoryear{{Henden} \& {Munari}}{{Henden} \&
  {Munari}}{2014}]{Henden14}
{Henden} A.,  {Munari} U.,  2014, Contributions of the Astronomical Observatory
  Skalnate Pleso, \href {http://adsabs.harvard.edu/abs/2014CoSka..43..518H}
  {43, 518}

\bibitem[\protect\citeauthoryear{{Henry}, {Marcy}, {Butler}  \& {Vogt}}{{Henry}
  et~al.}{2000}]{Henry00}
{Henry} G.~W.,  {Marcy} G.~W.,  {Butler} R.~P.,   {Vogt} S.~S.,  2000, \mn@doi
  [\apjl] {10.1086/312458}, \href
  {http://adsabs.harvard.edu/abs/2000ApJ...529L..41H} {529, L41}

\bibitem[\protect\citeauthoryear{{Irwin} et~al.,}{{Irwin}
  et~al.}{2004}]{Irwin04}
{Irwin} M.~J.,  et~al., 2004, in {Quinn} P.~J.,  {Bridger} A.,  eds,  \procspie
  Vol. 5493, Optimizing Scientific Return for Astronomy through Information
  Technologies. pp 411--422, \mn@doi{10.1117/12.551449}

\bibitem[\protect\citeauthoryear{{Irwin} et~al.,}{{Irwin}
  et~al.}{2011}]{Irwin11}
{Irwin} J.~M.,  et~al., 2011, \mn@doi [\apj] {10.1088/0004-637X/742/2/123},
  \href {http://adsabs.harvard.edu/abs/2011ApJ...742..123I} {742, 123}

\bibitem[\protect\citeauthoryear{{Jones} et~al.,}{{Jones}
  et~al.}{2016}]{Jones16}
{Jones} M.~I.,  et~al., 2016, \mn@doi [\aap] {10.1051/0004-6361/201628067},
  \href {http://adsabs.harvard.edu/abs/2016A%26A...590A..38J} {590, A38}

\bibitem[\protect\citeauthoryear{{Kaufer}, {Stahl}, {Tubbesing},
  {N{\o}rregaard}, {Avila}, {Francois}, {Pasquini}  \& {Pizzella}}{{Kaufer}
  et~al.}{1999}]{Kaufer99}
{Kaufer} A.,  {Stahl} O.,  {Tubbesing} S.,  {N{\o}rregaard} P.,  {Avila} G.,
  {Francois} P.,  {Pasquini} L.,   {Pizzella} A.,  1999, The Messenger, \href
  {http://adsabs.harvard.edu/abs/1999Msngr..95....8K} {95, 8}

\bibitem[\protect\citeauthoryear{{Kipping}}{{Kipping}}{2013}]{Kipping13}
{Kipping} D.~M.,  2013, \mn@doi [\mnras] {10.1093/mnras/stt1435}, \href
  {http://adsabs.harvard.edu/abs/2013MNRAS.435.2152K} {435, 2152}

\bibitem[\protect\citeauthoryear{{Kirk}, {Wheatley}, {Louden}, {Littlefair},
  {Copperwheat}, {Armstrong}, {Marsh}  \& {Dhillon}}{{Kirk}
  et~al.}{2016}]{Kirk16}
{Kirk} J.,  {Wheatley} P.~J.,  {Louden} T.,  {Littlefair} S.~P.,  {Copperwheat}
  C.~M.,  {Armstrong} D.~J.,  {Marsh} T.~R.,   {Dhillon} V.~S.,  2016, \mn@doi
  [\mnras] {10.1093/mnras/stw2205}, \href
  {http://adsabs.harvard.edu/abs/2016MNRAS.463.2922K} {463, 2922}

\bibitem[\protect\citeauthoryear{{Knutson} et~al.,}{{Knutson}
  et~al.}{2007}]{Knutson07}
{Knutson} H.~A.,  et~al., 2007, \mn@doi [\nat] {10.1038/nature05782}, \href
  {http://adsabs.harvard.edu/abs/2007Natur.447..183K} {447, 183}

\bibitem[\protect\citeauthoryear{{Koch} et~al.,}{{Koch} et~al.}{2010}]{Koch10}
{Koch} D.~G.,  et~al., 2010, \mn@doi [\apjl] {10.1088/2041-8205/713/2/L79},
  \href {http://adsabs.harvard.edu/abs/2010ApJ...713L..79K} {713, L79}

\bibitem[\protect\citeauthoryear{{Kov{\'a}cs}, {Zucker}  \&
  {Mazeh}}{{Kov{\'a}cs} et~al.}{2002}]{Kovacs02}
{Kov{\'a}cs} G.,  {Zucker} S.,   {Mazeh} T.,  2002, \mn@doi [\aap]
  {10.1051/0004-6361:20020802}, \href
  {http://adsabs.harvard.edu/abs/2002A%26A...391..369K} {391, 369}

\bibitem[\protect\citeauthoryear{{Kov{\'a}cs}, {Bakos}  \&
  {Noyes}}{{Kov{\'a}cs} et~al.}{2005}]{Kovacs05}
{Kov{\'a}cs} G.,  {Bakos} G.,   {Noyes} R.~W.,  2005, \mn@doi [\mnras]
  {10.1111/j.1365-2966.2004.08479.x}, \href
  {http://adsabs.harvard.edu/abs/2005MNRAS.356..557K} {356, 557}

\bibitem[\protect\citeauthoryear{{Kreidberg}}{{Kreidberg}}{2015}]{Kreidberg2015}
{Kreidberg} L.,  2015, \mn@doi [\pasp] {10.1086/683602}, \href
  {http://adsabs.harvard.edu/abs/2015PASP..127.1161K} {127, 1161}

\bibitem[\protect\citeauthoryear{{Kunder} et~al.,}{{Kunder}
  et~al.}{2017}]{Kunder17}
{Kunder} A.,  et~al., 2017, \mn@doi [\aj] {10.3847/1538-3881/153/2/75}, \href
  {http://adsabs.harvard.edu/abs/2017AJ....153...75K} {153, 75}

\bibitem[\protect\citeauthoryear{{Lang}, {Hogg}, {Mierle}, {Blanton}  \&
  {Roweis}}{{Lang} et~al.}{2010}]{Lang10}
{Lang} D.,  {Hogg} D.~W.,  {Mierle} K.,  {Blanton} M.,   {Roweis} S.,  2010,
  \mn@doi [\aj] {10.1088/0004-6256/139/5/1782}, \href
  {http://adsabs.harvard.edu/abs/2010AJ....139.1782L} {139, 1782}

\bibitem[\protect\citeauthoryear{{L{\'e}ger} et~al.,}{{L{\'e}ger}
  et~al.}{2009}]{Leger09}
{L{\'e}ger} A.,  et~al., 2009, \mn@doi [\aap] {10.1051/0004-6361/200911933},
  \href {http://adsabs.harvard.edu/abs/2009A%26A...506..287L} {506, 287}

\bibitem[\protect\citeauthoryear{{Lissauer} et~al.,}{{Lissauer}
  et~al.}{2011}]{Lissauer11}
{Lissauer} J.~J.,  et~al., 2011, \mn@doi [\nat] {10.1038/nature09760}, \href
  {http://adsabs.harvard.edu/abs/2011Natur.470...53L} {470, 53}

\bibitem[\protect\citeauthoryear{{Lissauer} et~al.,}{{Lissauer}
  et~al.}{2013}]{Lissauer13}
{Lissauer} J.~J.,  et~al., 2013, \mn@doi [\apj] {10.1088/0004-637X/770/2/131},
  \href {http://adsabs.harvard.edu/abs/2013ApJ...770..131L} {770, 131}

\bibitem[\protect\citeauthoryear{{Louden} \& {Wheatley}}{{Louden} \&
  {Wheatley}}{2015}]{Louden15}
{Louden} T.,  {Wheatley} P.~J.,  2015, \mn@doi [\apjl]
  {10.1088/2041-8205/814/2/L24}, \href
  {http://adsabs.harvard.edu/abs/2015ApJ...814L..24L} {814, L24}

\bibitem[\protect\citeauthoryear{{Mandel} \& {Agol}}{{Mandel} \&
  {Agol}}{2002}]{Mandel02}
{Mandel} K.,  {Agol} E.,  2002, \mn@doi [\apjl] {10.1086/345520}, \href
  {http://adsabs.harvard.edu/abs/2002ApJ...580L.171M} {580, L171}

\bibitem[\protect\citeauthoryear{{Mann}, {Feiden}, {Gaidos}, {Boyajian}  \&
  {von Braun}}{{Mann} et~al.}{2015}]{Mann15}
{Mann} A.~W.,  {Feiden} G.~A.,  {Gaidos} E.,  {Boyajian} T.,   {von Braun} K.,
  2015, \mn@doi [\apj] {10.1088/0004-637X/804/1/64}, \href
  {http://adsabs.harvard.edu/abs/2015ApJ...804...64M} {804, 64}

\bibitem[\protect\citeauthoryear{{Marmier} et~al.,}{{Marmier}
  et~al.}{2013}]{Marmier13}
{Marmier} M.,  et~al., 2013, \mn@doi [\aap] {10.1051/0004-6361/201219639},
  \href {http://adsabs.harvard.edu/abs/2013A%26A...551A..90M} {551, A90}

\bibitem[\protect\citeauthoryear{{Martin} et~al.,}{{Martin}
  et~al.}{2005}]{Martin05}
{Martin} D.~C.,  et~al., 2005, \mn@doi [\apjl] {10.1086/426387}, \href
  {http://adsabs.harvard.edu/abs/2005ApJ...619L...1M} {619, L1}

\bibitem[\protect\citeauthoryear{{Mayor} et~al.,}{{Mayor}
  et~al.}{2003}]{Mayor03}
{Mayor} M.,  et~al., 2003, The Messenger, \href
  {http://adsabs.harvard.edu/abs/2003Msngr.114...20M} {114, 20}

\bibitem[\protect\citeauthoryear{{McCauliff} et~al.,}{{McCauliff}
  et~al.}{2015}]{McCauliff15a}
{McCauliff} S.~D.,  et~al., 2015, \mn@doi [\apj] {10.1088/0004-637X/806/1/6},
  \href {http://adsabs.harvard.edu/abs/2015ApJ...806....6M} {806, 6}

\bibitem[\protect\citeauthoryear{{McCormac}, {Pollacco}, {Skillen}, {Faedi},
  {Todd}  \& {Watson}}{{McCormac} et~al.}{2013}]{McCormac13}
{McCormac} J.,  {Pollacco} D.,  {Skillen} I.,  {Faedi} F.,  {Todd} I.,
  {Watson} C.~A.,  2013, \mn@doi [\pasp] {10.1086/670940}, \href
  {http://adsabs.harvard.edu/abs/2013PASP..125..548M} {125, 548}

\bibitem[\protect\citeauthoryear{{McCormac} et~al.,}{{McCormac}
  et~al.}{2017}]{McCormac17}
{McCormac} J.,  et~al., 2017, \mn@doi [\pasp]
  {10.1088/1538-3873/129/972/025002}, \href
  {http://adsabs.harvard.edu/abs/2017PASP..129b5002M} {129, 025002}

\bibitem[\protect\citeauthoryear{{Motalebi} et~al.,}{{Motalebi}
  et~al.}{2015}]{Motalebi15}
{Motalebi} F.,  et~al., 2015, \mn@doi [\aap] {10.1051/0004-6361/201526822},
  \href {http://adsabs.harvard.edu/abs/2015A%26A...584A..72M} {584, A72}

\bibitem[\protect\citeauthoryear{{Moutou} et~al.,}{{Moutou}
  et~al.}{2005}]{Moutou05}
{Moutou} C.,  et~al., 2005, \mn@doi [\aap] {10.1051/0004-6361:20042334}, \href
  {http://adsabs.harvard.edu/abs/2005A%26A...437..355M} {437, 355}

\bibitem[\protect\citeauthoryear{{Moutou} et~al.,}{{Moutou}
  et~al.}{2007}]{Moutou07}
{Moutou} C.,  et~al., 2007, in {Afonso} C.,  {Weldrake} D.,   {Henning} T.,
  eds,  Astronomical Society of the Pacific Conference Series Vol. 366,
  Transiting Extrapolar Planets Workshop. p.~127

\bibitem[\protect\citeauthoryear{{Nikolov}, {Henning}, {Koppenhoefer}, {Lendl},
  {Maciejewski}  \& {Greiner}}{{Nikolov} et~al.}{2012}]{Nikolov12}
{Nikolov} N.,  {Henning} T.,  {Koppenhoefer} J.,  {Lendl} M.,  {Maciejewski}
  G.,   {Greiner} J.,  2012, \mn@doi [\aap] {10.1051/0004-6361/201118336},
  \href {http://adsabs.harvard.edu/abs/2012A%26A...539A.159N} {539, A159}

\bibitem[\protect\citeauthoryear{{Noll}, {Kausch}, {Barden}, {Jones},
  {Szyszka}, {Kimeswenger}  \& {Vinther}}{{Noll} et~al.}{2012}]{Noll12}
{Noll} S.,  {Kausch} W.,  {Barden} M.,  {Jones} A.~M.,  {Szyszka} C.,
  {Kimeswenger} S.,   {Vinther} J.,  2012, \mn@doi [\aap]
  {10.1051/0004-6361/201219040}, \href
  {http://adsabs.harvard.edu/abs/2012A%26A...543A..92N} {543, A92}

\bibitem[\protect\citeauthoryear{{Osborn}, {F{\"o}hring}, {Dhillon}  \&
  {Wilson}}{{Osborn} et~al.}{2015}]{Osborn15}
{Osborn} J.,  {F{\"o}hring} D.,  {Dhillon} V.~S.,   {Wilson} R.~W.,  2015,
  \mn@doi [\mnras] {10.1093/mnras/stv1400}, \href
  {http://adsabs.harvard.edu/abs/2015MNRAS.452.1707O} {452, 1707}

\bibitem[\protect\citeauthoryear{{Parviainen} \& {Aigrain}}{{Parviainen} \&
  {Aigrain}}{2015}]{Parviainen15}
{Parviainen} H.,  {Aigrain} S.,  2015, \mn@doi [\mnras]
  {10.1093/mnras/stv1857}, \href
  {http://adsabs.harvard.edu/abs/2015MNRAS.453.3821P} {453, 3821}

\bibitem[\protect\citeauthoryear{{Pecaut} \& {Mamajek}}{{Pecaut} \&
  {Mamajek}}{2013}]{Pecaut13}
{Pecaut} M.~J.,  {Mamajek} E.~E.,  2013, \mn@doi [\apjs]
  {10.1088/0067-0049/208/1/9}, \href
  {http://adsabs.harvard.edu/abs/2013ApJS..208....9P} {208, 9}

\bibitem[\protect\citeauthoryear{{Pepe} et~al.,}{{Pepe} et~al.}{2000}]{pepe00}
{Pepe} F.,  et~al., 2000, in {Iye} M.,  {Moorwood} A.~F.,  eds,  \procspie Vol.
  4008, Optical and IR Telescope Instrumentation and Detectors. pp 582--592

\bibitem[\protect\citeauthoryear{{Pepe} et~al.,}{{Pepe} et~al.}{2014}]{Pepe14}
{Pepe} F.,  et~al., 2014, \mn@doi [Astronomische Nachrichten]
  {12.1002/asna.201312004}, \href
  {http://adsabs.harvard.edu/abs/2014AN....335....8P} {335, 8}

\bibitem[\protect\citeauthoryear{{Pepper} et~al.,}{{Pepper}
  et~al.}{2017}]{Pepper17}
{Pepper} J.,  et~al., 2017, \mn@doi [\aj] {10.3847/1538-3881/aa62ab}, \href
  {http://adsabs.harvard.edu/abs/2017AJ....153..177P} {153, 177}

\bibitem[\protect\citeauthoryear{{Perryman}, {Hartman}, {Bakos}  \&
  {Lindegren}}{{Perryman} et~al.}{2014}]{Perryman14}
{Perryman} M.,  {Hartman} J.,  {Bakos} G.~{\'A}.,   {Lindegren} L.,  2014,
  \mn@doi [\apj] {10.1088/0004-637X/797/1/14}, \href
  {http://adsabs.harvard.edu/abs/2014ApJ...797...14P} {797, 14}

\bibitem[\protect\citeauthoryear{{Pickles}}{{Pickles}}{1998}]{Pickles98}
{Pickles} A.~J.,  1998, \mn@doi [\pasp] {10.1086/316197}, \href
  {http://adsabs.harvard.edu/abs/1998PASP..110..863P} {110, 863}

\bibitem[\protect\citeauthoryear{{Pollacco} et~al.,}{{Pollacco}
  et~al.}{2006}]{Pollacco06}
{Pollacco} D.~L.,  et~al., 2006, \mn@doi [\pasp] {10.1086/508556}, \href
  {http://adsabs.harvard.edu/abs/2006PASP..118.1407P} {118, 1407}

\bibitem[\protect\citeauthoryear{{Queloz} et~al.,}{{Queloz}
  et~al.}{2000}]{Queloz00}
{Queloz} D.,  et~al., 2000, \aap, \href
  {http://adsabs.harvard.edu/abs/2000A%26A...354...99Q} {354, 99}

\bibitem[\protect\citeauthoryear{{Queloz} et~al.,}{{Queloz}
  et~al.}{2009}]{Queloz09}
{Queloz} D.,  et~al., 2009, \mn@doi [\aap] {10.1051/0004-6361/200913096}, \href
  {http://adsabs.harvard.edu/abs/2009A%26A...506..303Q} {506, 303}

\bibitem[\protect\citeauthoryear{{Rauer} et~al.,}{{Rauer}
  et~al.}{2014}]{Rauer14}
{Rauer} H.,  et~al., 2014, \mn@doi [Experimental Astronomy]
  {10.1007/s10686-014-9383-4}, \href
  {http://adsabs.harvard.edu/abs/2014ExA....38..249R} {38, 249}

\bibitem[\protect\citeauthoryear{{Ricker} et~al.,}{{Ricker}
  et~al.}{2015}]{Ricker15}
{Ricker} G.~R.,  et~al., 2015, \mn@doi [Journal of Astronomical Telescopes,
  Instruments, and Systems] {10.1117/1.JATIS.1.1.014003}, \href
  {http://adsabs.harvard.edu/abs/2015JATIS...1a4003R} {1, 014003}

\bibitem[\protect\citeauthoryear{{Robin}, {Reyl{\'e}}, {Derri{\`e}re}  \&
  {Picaud}}{{Robin} et~al.}{2003}]{robin03}
{Robin} A.~C.,  {Reyl{\'e}} C.,  {Derri{\`e}re} S.,   {Picaud} S.,  2003,
  \mn@doi [\aap] {10.1051/0004-6361:20031117}, \href
  {http://adsabs.harvard.edu/abs/2003A%26A...409..523R} {409, 523}

\bibitem[\protect\citeauthoryear{{Roeser}, {Demleitner}  \&
  {Schilbach}}{{Roeser} et~al.}{2010}]{Roeser10}
{Roeser} S.,  {Demleitner} M.,   {Schilbach} E.,  2010, \mn@doi [\aj]
  {10.1088/0004-6256/139/6/2440}, \href
  {http://adsabs.harvard.edu/abs/2010AJ....139.2440R} {139, 2440}

\bibitem[\protect\citeauthoryear{{Schlegel}, {Finkbeiner}  \&
  {Davis}}{{Schlegel} et~al.}{1998}]{sfd98}
{Schlegel} D.~J.,  {Finkbeiner} D.~P.,   {Davis} M.,  1998, \mn@doi [\apj]
  {10.1086/305772}, \href {http://adsabs.harvard.edu/abs/1998ApJ...500..525S}
  {500, 525}

\bibitem[\protect\citeauthoryear{{Seager}, {Kuchner}, {Hier-Majumder}  \&
  {Militzer}}{{Seager} et~al.}{2007}]{Seager07}
{Seager} S.,  {Kuchner} M.,  {Hier-Majumder} C.~A.,   {Militzer} B.,  2007,
  \mn@doi [\apj] {10.1086/521346}, \href
  {http://adsabs.harvard.edu/abs/2007ApJ...669.1279S} {669, 1279}

\bibitem[\protect\citeauthoryear{{Sing} \& {L{\'o}pez-Morales}}{{Sing} \&
  {L{\'o}pez-Morales}}{2009}]{Sing09}
{Sing} D.~K.,  {L{\'o}pez-Morales} M.,  2009, \mn@doi [\aap]
  {10.1051/0004-6361:200811268}, \href
  {http://adsabs.harvard.edu/abs/2009A%26A...493L..31S} {493, L31}

\bibitem[\protect\citeauthoryear{{Sing} et~al.,}{{Sing} et~al.}{2016}]{Sing16}
{Sing} D.~K.,  et~al., 2016, \mn@doi [\nat] {10.1038/nature16068}, \href
  {http://adsabs.harvard.edu/abs/2016Natur.529...59S} {529, 59}

\bibitem[\protect\citeauthoryear{{Skrutskie} et~al.,}{{Skrutskie}
  et~al.}{2006}]{Skurtskie06}
{Skrutskie} M.~F.,  et~al., 2006, \mn@doi [\aj] {10.1086/498708}, \href
  {http://adsabs.harvard.edu/abs/2006AJ....131.1163S} {131, 1163}

\bibitem[\protect\citeauthoryear{{Snellen}, {de Kok}, {de Mooij}  \&
  {Albrecht}}{{Snellen} et~al.}{2010}]{Snellen10}
{Snellen} I.~A.~G.,  {de Kok} R.~J.,  {de Mooij} E.~J.~W.,   {Albrecht} S.,
  2010, \mn@doi [\nat] {10.1038/nature09111}, \href
  {http://adsabs.harvard.edu/abs/2010Natur.465.1049S} {465, 1049}

\bibitem[\protect\citeauthoryear{{Soto}, {Jenkins}  \& {Jones}}{{Soto}
  et~al.}{2015}]{Soto15}
{Soto} M.~G.,  {Jenkins} J.~S.,   {Jones} M.~I.,  2015, \mn@doi [\mnras]
  {10.1093/mnras/stv1144}, \href
  {http://adsabs.harvard.edu/abs/2015MNRAS.451.3131S} {451, 3131}

\bibitem[\protect\citeauthoryear{{Southworth} et~al.,}{{Southworth}
  et~al.}{2009}]{Southworth09}
{Southworth} J.,  et~al., 2009, \mn@doi [\mnras]
  {10.1111/j.1365-2966.2009.14767.x}, \href
  {http://adsabs.harvard.edu/abs/2009MNRAS.396.1023S} {396, 1023}

\bibitem[\protect\citeauthoryear{{Spiegel} \& {Burrows}}{{Spiegel} \&
  {Burrows}}{2013}]{Spiegel13}
{Spiegel} D.~S.,  {Burrows} A.,  2013, \mn@doi [\apj]
  {10.1088/0004-637X/772/1/76}, \href
  {http://adsabs.harvard.edu/abs/2013ApJ...772...76S} {772, 76}

\bibitem[\protect\citeauthoryear{{Surma}}{{Surma}}{1993}]{Surma1993}
{Surma} P.,  1993, \aap, \href
  {http://adsabs.harvard.edu/abs/1993A%26A...278..654S} {278, 654}

\bibitem[\protect\citeauthoryear{{Tamuz}, {Mazeh}  \& {Zucker}}{{Tamuz}
  et~al.}{2005}]{Tamuz05}
{Tamuz} O.,  {Mazeh} T.,   {Zucker} S.,  2005, \mn@doi [\mnras]
  {10.1111/j.1365-2966.2004.08585.x}, \href
  {http://adsabs.harvard.edu/abs/2005MNRAS.356.1466T} {356, 1466}

\bibitem[\protect\citeauthoryear{{Tingley}, {Bonomo}  \& {Deeg}}{{Tingley}
  et~al.}{2011}]{Tingley2011}
{Tingley} B.,  {Bonomo} A.~S.,   {Deeg} H.~J.,  2011, \mn@doi [\apj]
  {10.1088/0004-637X/726/2/112}, \href
  {http://adsabs.harvard.edu/abs/2011ApJ...726..112T} {726, 112}

\bibitem[\protect\citeauthoryear{{Triaud} et~al.,}{{Triaud}
  et~al.}{2010}]{Triaud10}
{Triaud} A.~H.~M.~J.,  et~al., 2010, \mn@doi [\aap]
  {10.1051/0004-6361/201014525}, \href
  {http://adsabs.harvard.edu/abs/2010A%26A...524A..25T} {524, A25}

\bibitem[\protect\citeauthoryear{{Uytterhoeven}, {Moya}, {Grigahc{\'e}ne},
  {Guzik}, { Guti{\'e}rrez-Soto}, {Smalley}  \& {Handler}}{{Uytterhoeven}
  et~al.}{2011}]{Uytterhoeven11}
{Uytterhoeven} K.,  {Moya} A.,  {Grigahc{\'e}ne} A.,  {Guzik} J.~A.,  {
  Guti{\'e}rrez-Soto} J.,  {Smalley} B.,   {Handler} J.,  2011, \mn@doi [\aap]
  {10.1051/0004-6361/201117368}, \href
  {http://adsabs.harvard.edu/abs/2011A%26A...534A.125U} {534, A125}

\bibitem[\protect\citeauthoryear{{Vidal-Madjar}, {Lecavelier des Etangs},
  {D{\'e}sert}, {Ballester}, {Ferlet}, {H{\'e}brard}  \&
  {Mayor}}{{Vidal-Madjar} et~al.}{2003}]{Vidal03}
{Vidal-Madjar} A.,  {Lecavelier des Etangs} A.,  {D{\'e}sert} J.-M.,
  {Ballester} G.~E.,  {Ferlet} R.,  {H{\'e}brard} G.,   {Mayor} M.,  2003,
  \mn@doi [\nat] {10.1038/nature01448}, \href
  {http://adsabs.harvard.edu/abs/2003Natur.422..143V} {422, 143}

\bibitem[\protect\citeauthoryear{Walker}{Walker}{2013}]{Walker13}
Walker S.,  2013, PhD thesis, University of Warwick, Coventry, UK

\bibitem[\protect\citeauthoryear{{Welsh} et~al.,}{{Welsh}
  et~al.}{2012}]{Welsh12}
{Welsh} W.~F.,  et~al., 2012, \mn@doi [\nat] {10.1038/nature10768}, \href
  {http://adsabs.harvard.edu/abs/2012Natur.481..475W} {481, 475}

\bibitem[\protect\citeauthoryear{{Wheatley} et~al.,}{{Wheatley}
  et~al.}{2013}]{Wheatley13}
{Wheatley} P.~J.,  et~al., 2013, in European Physical Journal Web of
  Conferences. p. 13002 (\mn@eprint {arXiv} {1302.6592}),
  \mn@doi{10.1051/epjconf/20134713002}

\bibitem[\protect\citeauthoryear{{Wilson} et~al.,}{{Wilson}
  et~al.}{2008}]{Wilson08}
{Wilson} D.~M.,  et~al., 2008, \mn@doi [\apjl] {10.1086/586735}, \href
  {http://adsabs.harvard.edu/abs/2008ApJ...675L.113W} {675, L113}

\bibitem[\protect\citeauthoryear{{Winn} et~al.,}{{Winn} et~al.}{2005}]{Winn05}
{Winn} J.~N.,  et~al., 2005, \mn@doi [\apj] {10.1086/432571}, \href
  {http://adsabs.harvard.edu/abs/2005ApJ...631.1215W} {631, 1215}

\bibitem[\protect\citeauthoryear{{Winn}, {Holman}, {Carter}, {Torres}, {Osip}
  \& {Beatty}}{{Winn} et~al.}{2009}]{Winn09}
{Winn} J.~N.,  {Holman} M.~J.,  {Carter} J.~A.,  {Torres} G.,  {Osip} D.~J.,
  {Beatty} T.,  2009, \mn@doi [\aj] {10.1088/0004-6256/137/4/3826}, \href
  {http://adsabs.harvard.edu/abs/2009AJ....137.3826W} {137, 3826}

\bibitem[\protect\citeauthoryear{{Winn} et~al.,}{{Winn} et~al.}{2011}]{Winn11}
{Winn} J.~N.,  et~al., 2011, \mn@doi [\apjl] {10.1088/2041-8205/737/1/L18},
  \href {http://adsabs.harvard.edu/abs/2011ApJ...737L..18W} {737, L18}

\bibitem[\protect\citeauthoryear{{Young}}{{Young}}{1967}]{Young67}
{Young} A.~T.,  1967, \mn@doi [\aj] {10.1086/110303}, \href
  {http://adsabs.harvard.edu/abs/1967AJ.....72..747Y} {72, 747}

\bibitem[\protect\citeauthoryear{{Zacharias}, {Finch}, {Girard}, {Henden},
  {Bartlett}, {Monet}  \& {Zacharias}}{{Zacharias} et~al.}{2013}]{Zacharias13}
{Zacharias} N.,  {Finch} C.~T.,  {Girard} T.~M.,  {Henden} A.,  {Bartlett}
  J.~L.,  {Monet} D.~G.,   {Zacharias} M.~I.,  2013, \mn@doi [\aj]
  {10.1088/0004-6256/145/2/44}, \href
  {http://adsabs.harvard.edu/abs/2013AJ....145...44Z} {145, 44}

\bibitem[\protect\citeauthoryear{{Zissell}}{{Zissell}}{2000}]{Zissell00}
{Zissell} R.~E.,  2000, Journal of the American Association of Variable Star
  Observers (JAAVSO), \href {http://adsabs.harvard.edu/abs/2000JAVSO..28..149Z}
  {28, 149}

\makeatother
\end{thebibliography}








\bsp	
\label{lastpage}
\end{document}